\documentclass[aps,pre,reprint,groupedaddress]{revtex4-2}

\usepackage{graphicx}
\usepackage{amsmath,amssymb}
\begin{document}

%\title{Extracting Dynamical Frequencies from Invariants of Motion in Accelerator Designs Based on Coupled Nonlinear Integrable Systems}
\title{Extracting Dynamical Frequencies from Invariants of Motion in Finite-Dimensional Nonlinear Integrable Systems}
%\title{Obtaining Frequencies from Invariants of Motion in Coupled Nonlinear Integrable Dynamical Systems}

\author{Chad E. Mitchell}
\email{ChadMitchell@lbl.gov}
\author{Robert D. Ryne}
 \author{Kilean Hwang}
 
\address{Lawrence Berkeley National Laboratory, Berkeley, CA  94720, USA}

\author{Sergei Nagaitsev}
\altaffiliation[Also at ]{the University of Chicago,
Chicago, Illinois 60637, USA}
\author{Timofey Zolkin}

\address{Fermi National Accelerator Laboratory, Batavia, IL  60510, USA}

\date{\today}

\begin{abstract}
Integrable dynamical systems play an important role in many areas of science, including accelerator and plasma physics. An integrable dynamical system with $n$ degrees of freedom (DOF) possesses $n$ nontrivial integrals of motion, and can be solved, in principle, by covering the phase space with one or more charts in which the dynamics can be described using action-angle coordinates. To obtain the frequencies of motion, both the transformation to action-angle coordinates and its inverse must be known in explicit form.  However, no general algorithm exists for constructing this transformation explicitly from a set of $n$ known (and generally coupled) integrals of motion. 
In this paper we describe how one can determine the dynamical frequencies of the motion as functions of these $n$ integrals in the absence of explicitly-known action-angle variables, and we provide several examples.  
\end{abstract}

%\keywords{}

\maketitle

%\linenumbers


\section{Introduction}
Integrable dynamical systems play an important role in many areas of science, including accelerator \cite{Danilov,Nbody} and plasma physics.
It is well-known that an $n$-DOF integrable system can be solved, in principle, by constructing action-angle coordinates.   However, in general such action-angle coordinates are defined only locally, and break down near critical phase space structures (e.g., the separatrix of the nonlinear pendulum).  In addition, the canonical transformation to action-angle coordinates is difficult to obtain in explicit closed form for even the simplest systems. In practice, this can be an obstacle to extracting the dynamical frequencies of motion of the system, which are often the primary quantities of interest.  Finally, the trend in mechanics is to move toward results that can be expressed in a geometric form, independent of a specific choice of coordinates.

In this paper, we propose a method to find the $n$ dynamical frequencies of an integrable symplectic map or a Hamiltonian flow without knowledge of the transformation to action-angle coordinates.  This result is motivated by the Mineur-Arnold formula \cite{Bolsinov,Arnold,Marsden,Moser}, which states that the $n$ action coordinates $I_j$ can be constructed as path integrals of the form:
\begin{equation}
    I_j=\frac{1}{2\pi}\oint_{\gamma_j}\sum_{k=1}^np_k dq_k,\quad (j=1,\ldots,n), \label{Mineur}
\end{equation}
where the $\gamma_j$ define $n$ appropriately-chosen closed paths (cycles) in the invariant level set (Appendix A).  We will show that an explicit integral formula analogous to (\ref{Mineur}) can be obtained for the $n$ dynamical frequencies.  This result is a generalization to arbitrary dimension of a result described in \cite{Zolkin}, which is valid for the special case when $n=1$.

It is emphasized that this procedure is developed for the narrow class of Hamiltonian systems (or symplectic maps) with a sufficient number of exactly-known invariants, and not for arbitrary Hamiltonian systems.  However, experience suggests that this procedure may be used to extract and to understand the frequency behavior of systems for which ``approximate invariants" can be constructed, which exhibit sufficiently small variation over the time scale of interest.  Such quantities can sometimes be constructed analytically or numerically \cite{MLinvariants1, MLinvariants2}.

The structure of this paper is as follows.  Section II provides a brief summary of background definitions regarding integrable maps and flows.  Section III motivates the concept of the tunes (or equivalently, the rotation vector) of an integrable symplectic map.  Section IV contains the main technical result of this paper (\ref{invariant}), relating the tunes of an integrable symplectic map to its dynamical invariants.  Section V describes the mathematical properties of this solution, together with its proof.  In Section VI, we describe how this result can be applied to determine the characteristic frequencies of an integrable Hamiltonian flow.  Section VII illustrates the application of these results using two numerical examples.  Conclusions are provided in Section VIII.  There are four Appendices.

\section{Integrable Maps and Flows}
For simplicity, we take the phase space $M$ to be an open subset of $\mathbb{R}^{2n}$ with its standard symplectic form.  In any local set of canonical coordinates $(q_1,\ldots,q_n,p_1,\ldots,p_n)$, the symplectic form is represented by the matrix:
\begin{equation}
J=\begin{pmatrix}
0 & I_{n\times n} \\
-I_{n\times n} & 0
\end{pmatrix}. \label{Jmat}
\end{equation}
We will frequently use the fact that $J^T=J^{-1}=-J$.

Let $\mathcal{M}:M\rightarrow M$ denote a symplectic map.  A smooth function $f:M\rightarrow\mathbb{R}$ is said to be an {\it invariant} of the map $\mathcal{M}$ if:
\begin{equation}
f\circ\mathcal{M}=f. \label{composite}
\end{equation}
The map $\mathcal{M}$ is said to be {\it completely integrable} if there exists a set of $n$ invariants $f_k$ such that:  i) the invariants Poisson-commute:  $\{f_j,f_k\}=0$ $(j,k=1,\ldots,n)$, and ii) the set of gradient vectors $\nabla f_k$ $(k=1,\ldots,n)$ is linearly independent at every point of $M$, except for a possible set of zero measure (phase space volume) \cite{integrableMaps1,integrableMaps2,integrableMaps3}.

Similarly, if $H:M\rightarrow\mathbb{R}$ denotes a smooth Hamiltonian function, the flow generated by $H$ is said to be completely integrable if the conditions i)-ii) apply, with
the invariant condition (\ref{composite}) replaced by the local condition $\{f,H\}=0$.

To analyze the behavior of such a map or a flow, let $\mathcal{F}:M\rightarrow\mathbb{R}^n$ denote the {\it momentum mapping}, the function that takes each point in the phase space to its $n$-tuple of invariants \cite{Bolsinov}:
\begin{equation}
\mathcal{F}(\zeta)=(f_1(\zeta),\ldots,f_n(\zeta)),\quad\quad\zeta\in M. \label{mommap}
\end{equation}
Each orbit is then confined to lie in some level set of $\mathcal{F}$ of the form:
\begin{equation}
M_c=\{\zeta\in M :\mathcal{F}(\zeta)=c\},\quad\quad c\in \mathbb{R}^n. \label{level}
\end{equation}
The level set (\ref{level}) is said to be {\it regular} if the linear map $D\mathcal{F}$, represented by the Jacobian matrix of $\mathcal{F}$, is surjective (rank $n$) everywhere on $M_c$.  In this case, $M_c$ is a smooth surface of dimension $n$. Assuming that $M_c$ is also compact and connected, the Liouville-Arnold theorem \cite{Bolsinov,Arnold,Marsden,Moser} states that $M_c$ may be smoothly transformed by a symplectic change of coordinates into the standard $n$-torus $\mathbb{T}^n$, and application of the map (or flow) corresponds to rotation about this torus with a fixed frequency vector, which we wish to determine.

\section{Tunes of an Integrable Map}
Let $\mathcal{M}$ be an integrable symplectic map, and let $M_c$ be one of its regular level sets.  
By the Liouville-Arnold theorem, there exists a neighborhood of the level set $M_c$ in which there is a set of canonical action-angle coordinates 
$\zeta=(\phi_1,\ldots,\phi_n,I_1,\ldots,I_n)$ in which the map takes the form $\mathcal{M}({\phi},{I})=({\phi}^f,{I}^f)$, where:
\begin{equation}
{I}^f={I},\quad\quad {\phi}^f={\phi}+2\pi\nu({I})\quad\operatorname{mod}2\pi. \label{tunemap0}
\end{equation} 
The coordinates $(\phi,I)$ in (\ref{tunemap0}) are not unique \cite{AA1}.  However, the quantities $\nu_j$ $(j=1,\ldots,n)$, called the {\it tunes} of $\mathcal{M}$, have a coordinate-invariant physical meaning, described as follows.  

If $F$ denotes any observable, given by a smooth real-valued function defined in our neighborhood of $M_c$, then $F$ may be expressed as a uniformly convergent Fourier series in the angle coordinates $\phi$, so that:
\begin{equation}
F(\phi,I)=\sum_{k\in \mathbb{Z}^n}a_k(I)e^{ik\cdot\phi},\quad\quad a_k\in\mathbb{C}.
\end{equation}
  Applying the map $\mathcal{M}$ in the form (\ref{tunemap0}) $N$ times shows that:
 \begin{equation}
 F(\mathcal{M}^N(\phi,I))=\sum_{k\in\mathbb{Z}^N}a_k(I)e^{ik\cdot(\phi+2\pi k\cdot\nu(I)N)}. \label{observable}
 \end{equation}
From (\ref{observable}), it follows that there exist smooth complex-valued functions $F_k$ $(k\in\mathbb{Z}^n)$ on our neighborhood of $M_c$ such that:
\begin{equation}
F\circ\mathcal{M}^N=\sum_{k\in\mathbb{Z}^n}F_{{k}}e^{i2\pi({k}\cdot{\nu})N}. \label{tunedef}
\end{equation}  

One sees from (\ref{tunedef}) that any time series obtained by following an observable $F$ (defined on the level set $M_c$) during iteration of the map $\mathcal{M}$ contains contributions at the discrete set of frequencies:
\begin{equation}
\Omega_{\nu}=\{k\cdot\nu+k_0:k\in\mathbb{Z}^n,k_0\in\mathbb{Z}\}. \label{freqset}
\end{equation}
Algorithms to determine the basic frequencies $\nu_j$ $(j=1,\ldots,n)$ from a  series of the form (\ref{tunedef}) are well-established \cite{Laskar1,Laskar2}.  

Note that knowledge of the set of frequencies (\ref{freqset}) does not specify the vector $\nu\in\mathbb{R}^n$ uniquely.  To see this, let
\begin{equation}
\nu'=U\nu+m, \label{unimodular}
\end{equation}
where $m\in\mathbb{Z}^n$ is any $n$-tuple of integers and $U$ is any unimodular integer matrix (an $n\times n$ integer matrix with $\operatorname{det}U=\pm 1$).  This implies that $U$ is invertible, $U^{-1}$ is also a unimodular integer matrix, and $U$ defines an invertible linear transformation from $\mathbb{Z}^n$ to $\mathbb{Z}^n$.  The same conclusion holds for $U^T$.  By making the transformation of integer indices $k=U^Tk'$, the sum in (\ref{tunedef}) becomes:
\begin{equation}
F\circ\mathcal{M}^N=\sum_{k'\in\mathbb{Z}^n}F_{U^T{k'}}e^{i2\pi({k'}\cdot{\nu'})N},
\end{equation}
which takes the same form as (\ref{tunedef}), with $\nu$ replaced by $\nu'$.  A similar argument starting from (\ref{freqset}) shows that $\Omega_{\nu'}=\Omega_{\nu}$.
Thus, the vector $\nu$ is at best defined only up to transformations of the form (\ref{unimodular}) \cite{Affine}.  

Indeed, one can construct action-angle coordinates in which the map $\mathcal{M}$ has the form (\ref{tunemap0}) with the tunes $\nu'$ given by (\ref{unimodular}).  In terms of the original coordinates $(\phi,I)$, let:
\begin{equation}
I'=U^{-T}I,\quad\quad \phi'=U\phi\quad\operatorname{mod}2\pi. \label{aa}
\end{equation}
The quantities $(\phi'_1,\ldots,\phi'_n)$ define periodic angle coordinates on the torus $\mathbb{T}^n$, since $\phi^A=\phi^B$ $\operatorname{mod}2\pi\Leftrightarrow U\phi^A=U\phi^B$ $\operatorname{mod}2\pi$, by the condition that $U$ be a unimodular integer matrix.  The transformation (\ref{aa}) is easily verified to be symplectic.  The map $\mathcal{M}$ in the coordinates $(\phi',I')$ takes the form:
\begin{equation}
I'^f=I',\quad\quad \phi'^f=\phi'+2\pi\nu'(I')\quad \operatorname{mod}2\pi,
\end{equation}
where
\begin{equation}
\nu'(I')=U\nu(U^TI')+m.
\end{equation}
Since points on the level set $M_c$ satisfy a condition of the form $I_0=I=U^TI'$ for some constant $I_0\in\mathbb{R}^n$, it follows that (\ref{unimodular}) holds on $M_c$, as claimed.

The vector $\nu$ is called the {\it rotation vector} of the map $\mathcal{M}$ corresponding to the level set $M_c$ \cite{Katok}.  Two rotation vectors $\nu$ and $\nu'$ will be said to be {\it equivalent} if there exists a relation of the form (\ref{unimodular}).  In practice, one would like a natural method to select a unique representative from each equivalence class.  In addition, one would like the selected vector $\nu$ to vary smoothly with the invariant value $c\in\mathbb{R}^n$. If the map $\mathcal{M}$ decouples when expressed using a particular choice of canonical coordinates, then the $n$ tunes can be chosen (up to a permutation) to correspond to rotation angles in each of the $n$ conjugate phase planes.  If the system is coupled, then selecting a natural choice of representative is a more subtle issue.  However, note that the rotation vector $\nu$ may always be chosen so that $0\leq \nu_j \leq 1/2$ $(j=1,\ldots,n)$.

The precise choice of the rotation vector is closely related to geometric considerations.  In the following section, we will see that there is a correspondence between the rotation vector and the choice of certain paths lying in the invariant torus.  It is of interest to study the relationships between the analytic properties of the rotation vector and the topology of these curves.  However, for the remainder of this paper, we content ourselves with demonstrating that all results are valid up to an equivalance of the form (\ref{unimodular}). 

\section{Tunes from invariants}
Let $\mathcal{M}$ be an integrable symplectic map with momentum mapping $\mathcal{F}$, as defined in (\ref{mommap}).  
The goal of this paper is to demonstrate that on any regular level set of $\mathcal{F}$, the tunes $\nu=(\nu_1,\ldots,\nu_n)^T$ may be expressed using a set of  $n(n+1)$ path integrals over the level set, in the form:
\begin{subequations}\label{invariant}
\begin{align}
{S}&=-\int_{\gamma}(D\mathcal{F}^+)^TJd{\zeta}, \\ \quad \label{invariantform}
R_{j k}&=\left(-\oint_{\gamma_k}(D\mathcal{F}^+)^TJd{\zeta}\right)_j , \\
{\nu}&=R^{-1}{S}. \label{invarianttunes}
\end{align}
\end{subequations}

Here $\nu$ and $S$ are real $n$-vectors, $R$ is a real $n\times n$ matrix, and $J$ is the $2n\times 2n$ matrix of the symplectic form (\ref{Jmat}).  It will be shown that the matrix $R$ is, in fact, invertible.

In (\ref{invariant}), $\gamma$ is a parameterized path in the level set $M_c$ from any point $\zeta\in M_c$ to its image $\mathcal{M}(\zeta)$ under the map.  Likewise, the $\gamma_k$ ($k=1,\ldots,n$) are parameterized closed paths in the level set $M_c$, and these must be chosen to form a basis for the group of 1-cycles in $M_c$.  (See Appendix A.)  We will show that the resulting value of $\nu\in\mathbb{R}^n$ is independent, modulo the equivalence (\ref{unimodular}), of the choice of the paths $\gamma$ and $\gamma_k$.  Furthermore, the precise value of $\nu$ depends only on the topology of the curves $\gamma$ and $\gamma_k$.  Intuitively, the paths $(\gamma_1,\ldots,\gamma_n)$ specify $n$ independent ``winding directions" around the level set $M_c$, and the tunes $(\nu_1,\ldots,\nu_n)$ specify the fraction of a cycle (in each direction) by which a point is moved under application the map $\mathcal{M}$.

Finally, $D\mathcal{F}^+$ denotes any $2n\times n$ right matrix inverse of the $n\times 2n$ Jacobian matrix $D\mathcal{F}$.  Since $\operatorname{rank}(D\mathcal{F})=n$ on the level set $M_c$, such a right inverse exists at every point on $M_c$.  It is convenient to use the Moore-Penrose inverse of $D\mathcal{F}$, given explicitly by:
\begin{equation}
D\mathcal{F}^+=(D\mathcal{F})^T\left[(D\mathcal{F})(D\mathcal{F})^T\right]^{-1}. \label{MooreP}
\end{equation}
By the rank assumption on $D\mathcal{F}$, the matrix appearing in square brackets in (\ref{MooreP}) is always invertible.  It follows that the matrix elements of $D\mathcal{F}^+$
are smooth, bounded functions when restricted to the level set $M_c$, and the path integrals in (\ref{invariant}) are convergent and finite.  Appendix B describes important properties of the matrix $D\mathcal{F}^+$ that are used in the remainder of this paper.

\subsection{Simple Example}
\begin{figure}
\begin{center}
\includegraphics[width=2.5in]{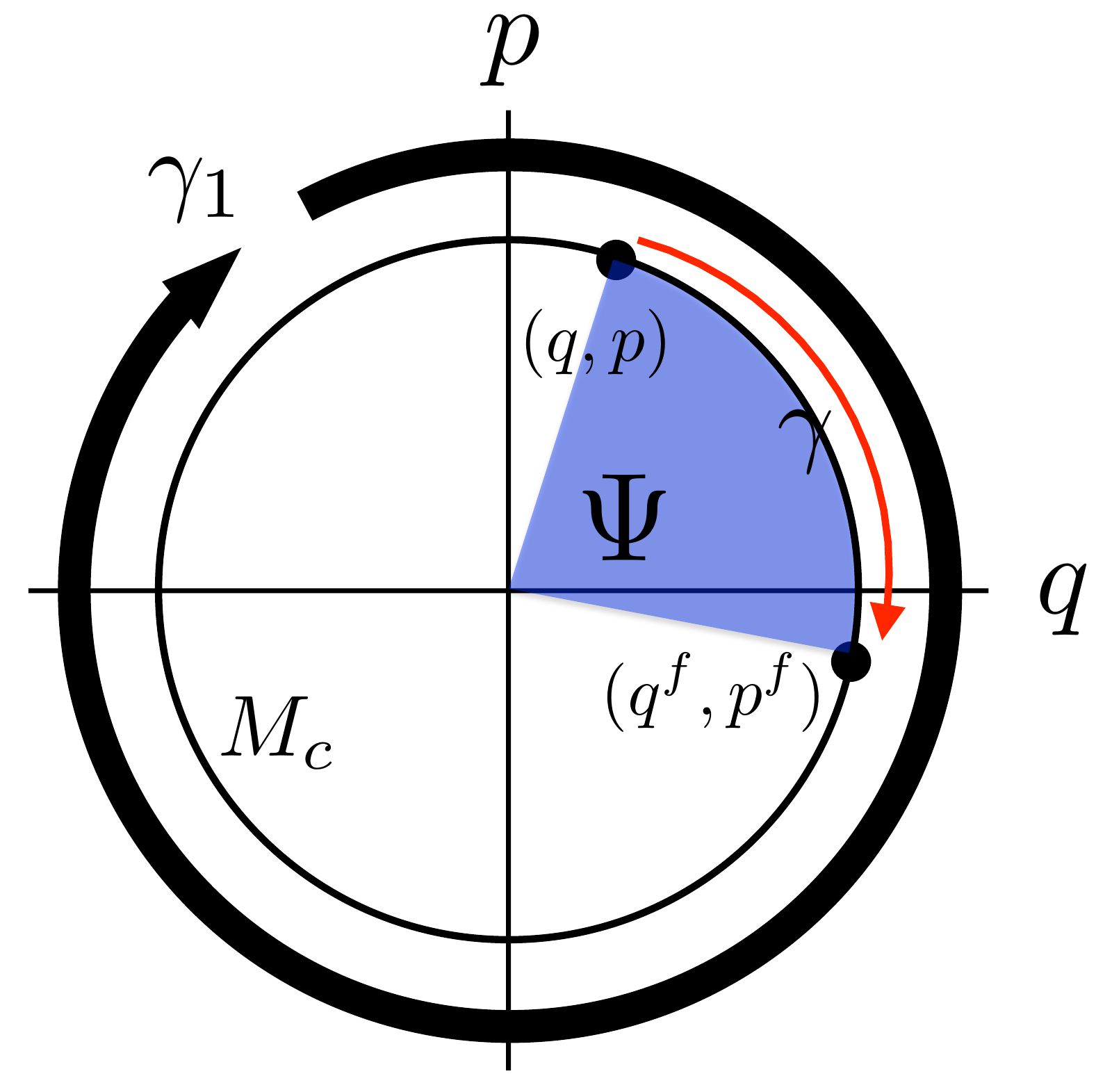}
\end{center}
\caption{ Illustration of the map (\ref{example0}), showing one of the level sets $M_c$ $(c>0)$ of the invariant $f$ in (\ref{fsimple}) and the two curves $\gamma$ (red) and $\gamma_1$ (black) used to evaluate (\ref{invariant}).  Although not shown here, each curve is allowed to change direction during transit.  The curve $\gamma$ may wind around the origin multiple times.}
\label{fig:Simple}
\end{figure}

Consider the 2D linear symplectic map described in matrix form as:
\begin{equation}
\begin{pmatrix}
q^f \\ p^f
\end{pmatrix}=
\begin{pmatrix}
\cos\Psi & \sin\Psi \\
-\sin\Psi & \cos\Psi
\end{pmatrix}
\begin{pmatrix}
q \\ p
\end{pmatrix}, \label{example0}
\end{equation}
which arises naturally in the study of the simple harmonic oscillator.
In this case $n=1$ and an invariant is given by:
\begin{equation}
f(q,p)=\frac{1}{2}(q^2+p^2).\label{fsimple}
\end{equation}
The level set $M_c=\{(q,p)\in\mathbb{R}^2:f(q,p)=c\}$ is regular for any $c>0$, corresponding to the circle of radius $\sqrt{2c}$ with center at the origin.  (See Fig. \ref{fig:Simple}.)
We therefore express the two curves $\gamma$ and $\gamma_1$ appearing in (\ref{invariant}) as:
\begin{subequations}\label{gammaparam}
\begin{align}
\gamma(t)=(\sqrt{2c}\cos\alpha(t),\sqrt{2c}\sin\alpha(t)),\quad a\leq t\leq b \\
\gamma_1(t)=(\sqrt{2c}\cos\beta(t),\sqrt{2c}\sin\beta(t)),\quad c\leq t\leq d
\end{align}
\end{subequations}
where $\alpha$ and $\beta$ are (smooth) real-valued functions of some parameter $t$.  
The definitions of $\gamma$ and $\gamma_1$ in (\ref{invariant}) require only that the functions $\alpha$ and $\beta$ satisfy:
\begin{equation}
\alpha(b)=\alpha(a)-\Psi-2\pi m, \quad \beta(d)=\beta(c)\mp 2\pi,
\end{equation}
where $m$ may be any integer.  (In order to serve as a basis cycle, the curve $\gamma_1$ must wind around the circle exactly once, in either direction.)
One verifies using (\ref{fsimple}) that, since $\mathcal{F}=f$ we have:
\begin{equation}
D\mathcal{F}=\begin{pmatrix}
q & p \end{pmatrix},\quad\quad 
D\mathcal{F}^+=\frac{1}{q^2+p^2}
\begin{pmatrix}
q \\ p \end{pmatrix}.
\end{equation}
Using these results in (\ref{invariant}) gives:
\begin{align}
S&=-\int_a^b(D\mathcal{F}^+)^TJ\gamma'(t)dt=-\int_a^b\alpha'(t)dt=\Psi+2\pi m, \notag \\
R&=-\int_c^d(D\mathcal{F}^+)^TJ\gamma_1'(t)dt=-\int_c^d\beta'(t)dt=\pm 2\pi. \notag
\end{align}
This yields the following result for the tune $\nu$ of the map (\ref{example0}):
\begin{equation}
\nu=R^{-1}S=\pm\left(\frac{\Psi}{2\pi}+ m\right),\quad\quad m\in\mathbb{Z}. \label{tune0}
\end{equation}

This result is expected, since (\ref{example0}) represents a clockwise rotation in the phase space by the angle $\Psi$.
If we think of the basis cycle $\gamma_1$ as defining an orientation of the circle $M_c$ (i.e., defining the clockwise or the counterclockwise direction to be positive), then $\nu$ represents the fraction of a cycle that is completed as we move along the curve $\gamma$, completing one iteration of the map.  The sign in (\ref{tune0}) is determined by the direction of $\gamma$, while the integer $m$ counts the number of complete turns that the curve $\gamma$ winds about the origin.
Note that the final result is independent of the parametrization (\ref{gammaparam}), as defined by the choices of the functions $\alpha$ and $\beta$.  

The purpose of this example is to illustrate the result (\ref{invariant}) in its simplest possible setting.
More sophisticated examples are considered in Section VII, in Appendices C-D, and in the reference \cite{Zolkin}.

\section{\label{sec:prop} Properties of the Solution}
In this section, we discuss the properties of the general solution (\ref{invariant}), and we provide its mathematical proof.
\subsection{\label{sec:path} Path integrals in the level set}
If $A:M\rightarrow\mathbb{R}^{n\times 2n}$ is a smooth matrix-valued function on the phase space, and if $\gamma:[a,b]\rightarrow M$ is a smooth parametrized path, then
an integral of the form (\ref{invariant}) is to be interpreted as:
\begin{equation}
\int_{\gamma}Ad\zeta=\int_a^bA(\gamma(t))\gamma'(t)dt, \label{pathdef}
\end{equation}
where $\gamma'(t)$ is the $2n$-vector tangent to $\gamma$ at $t$.
For any path $\gamma$ confined to a level set of $\mathcal{F}$, $\mathcal{F}$ is invariant along $\gamma$, and applying the chain rule gives that:
\begin{equation}
0=\frac{d}{dt}(\mathcal{F}\circ\gamma)(t)=D\mathcal{F}({\gamma(t)})\gamma'(t). \label{tangent0}
\end{equation}
Since this holds for every such path $\gamma$, motivated by (\ref{pathdef}) we will denote (\ref{tangent0}) more simply as:
\begin{equation}
(D\mathcal{F})d{\zeta}=0. \label{tangent}
\end{equation}
Since it follows from (\ref{tangent}) that $Jd{\zeta}\in J\operatorname{ker}(D\mathcal{F})$, we have from (\ref{project}) that:
\begin{equation}
(D\mathcal{F}^+)(D\mathcal{F})Jd{\zeta}=Jd{\zeta}.
\end{equation}
Since $(D\mathcal{F}^+)(D\mathcal{F})$ is symmetric, as is easily verified, we have:
\begin{equation}
(D\mathcal{F})^T(D\mathcal{F}^+)^TJd{\zeta}=Jd{\zeta}. \label{invid}
\end{equation}
The identity (\ref{invid}) allows us to prove many results on coordinate and path independence of the integrals in (\ref{invariant}).

As an example, let $B$ denote {\it any} right matrix inverse of $D\mathcal{F}$.  Then $B^T$ is a left inverse of $(D\mathcal{F})^T$.  Multiplying (\ref{invid}) on the left by $B^T$ gives:
\begin{equation}
(D\mathcal{F}^+)^TJd{\zeta}=B^TJd{\zeta},
\end{equation}
which shows that we could replace $D\mathcal{F}^+$ by any right matrix inverse of $D\mathcal{F}$ in the integrals (\ref{invariant}) without changing the result.

\subsection{Coordinate-independence}
Let $\zeta'$ denote a vector of new phase space coordinates related to $\zeta$ by an arbitrary symplectic coordinate transformation $\mathcal{N}$, so that
\begin{equation}
\zeta'=\mathcal{N}(\zeta).
\end{equation}
Let all quantities expressed in these new coordinates be denoted with $'$.  Then it is straightforward to verify that:
\begin{equation}
d\zeta'=(D\mathcal{N})d\zeta,\quad D\mathcal{F}'=(D\mathcal{F})(D\mathcal{N})^{-1}. \label{transf1}
\end{equation}
Since the map $\mathcal{N}$ is symplectic:
\begin{equation}
(D\mathcal{N})^{T}J(D\mathcal{N})=J. \label{transf2}
\end{equation}
To simplify notation, let $dv$ denote the form appearing in the integrals (\ref{invariant}), namely
\begin{equation}
dv=(D\mathcal{F}^+)^TJd\zeta. \label{vdef}
\end{equation}
Writing down the identity (\ref{invid}) in the primed coordinates, we have:
\begin{equation}
(D\mathcal{F'})^T{dv}'=Jd\zeta'. \label{start}
\end{equation}
Making the substitutions of (\ref{transf1}) into (\ref{start}) gives:
\begin{equation}
D\mathcal{N}^{-T}(D\mathcal{F})^Tdv'=J(D\mathcal{N})d\zeta.
\end{equation}
Multiplying both sides by $D\mathcal{N}^T$ gives
\begin{equation}
(D\mathcal{F})^Tdv'=(D\mathcal{N})^TJ(D\mathcal{N})d\zeta.
\end{equation}
Applying the symplectic condition (\ref{transf2}) gives:
\begin{equation}
(D\mathcal{F})^Tdv'=Jd\zeta.
\end{equation}

Finally, multiplying both sides by $(D\mathcal{F}^+)^T$ and noting that this is a left inverse of $(D\mathcal{F})^T$ gives:
\begin{equation}
dv'=(D\mathcal{F}^+)^TJd\zeta=dv. \label{coordind}
\end{equation}
Since (\ref{invariant}) can be written as:
\begin{equation}
S=-\int_{\gamma}dv,\quad R_{jk}=\left(-\oint_{\gamma_k}dv\right)_j, \label{coordind2}
\end{equation}
it follows from (\ref{coordind}) that for a fixed choice of paths $\gamma$ and $\gamma_k$ $(k=1,\ldots,n)$
each integral in (\ref{coordind2}) is independent of the choice of canonical coordinates.

\subsection{Reduced forms in canonical coordinates}
Consider canonical coordinates given by ${\zeta}=(q_1,\ldots,q_n,p_1,\ldots,p_n)^T$.  
We may express the $n\times 2n$ matrix $D\mathcal{F}$ in terms of two $n\times n$ blocks, which correspond to partial derivatives with respect to the variables $q=(q_1,\ldots,q_n)$ and $p=(p_1,\ldots,p_n)$, respectively: 
\begin{equation}
D\mathcal{F}=\begin{pmatrix}
D_q\mathcal{F} & D_p\mathcal{F}
\end{pmatrix}. \label{Fmat}
\end{equation}
Let $dv$ be defined as in (\ref{vdef}).  Then using identity (\ref{invid}) gives:
\begin{equation}
(D\mathcal{F})^Tdv=Jd{\zeta}.
\end{equation}
Expressing this in terms of its $n\times n$ blocks using (\ref{Jmat}) and (\ref{Fmat}) gives:
\begin{equation}
\begin{pmatrix}
D_q\mathcal{F}^Tdv \\
D_p\mathcal{F}^Tdv
\end{pmatrix} = 
\begin{pmatrix}
d{p} \\ -d{q}
\end{pmatrix}. \label{blockparts}
\end{equation}
In the special case that the matrix $(D_q\mathcal{F})^T$ is invertible along the integration path, we may use the first row in (\ref{blockparts}) to give:
\begin{equation}
dv=(D_q\mathcal{F})^{-T}d{p}.
\end{equation}
Noting the definition of $dv$ it follows that:
\begin{subequations}\label{TS0}
\begin{align}
{S}&=-\int_{\gamma}(D_q\mathcal{F})^{-T}d{p}, \\
R_{j k}&=\left(-\oint_{\gamma_k}(D_q\mathcal{F})^{-T}d{p}\right)_j,\quad {\nu}=R^{-1}{S}.
\end{align}
\end{subequations}

Alternatively, in the special case that the matrix $(D_p\mathcal{F})^T$ is invertible along the integration path, we may use the second row in (\ref{blockparts}) to give:
\begin{equation}
dv=-(D_p\mathcal{F})^{-T}d{q}.
\end{equation}
Noting the definition of $dv$ it follows that:
\begin{subequations} \label{TS}
\begin{align}
{S}&=\int_{\gamma}(D_p\mathcal{F})^{-T}d{q}, \\
R_{j k}&=\left(\oint_{\gamma_k}(D_p\mathcal{F})^{-T}d{q}\right)_j,\quad {\nu}=R^{-1}{S}.
\end{align}
\end{subequations}
In the special case of one degree of freedom $(n=1)$, the expression (\ref{TS}) reduces to the expression appearing in \cite{Zolkin}.  Another example, for a map with two degrees of freedom $(n=2)$ separable in polar coordinates, is provided in Appendix D.

\subsection{Proof of the result}\label{Proof}
By the Liouville-Arnold theorem for integrable symplectic maps, there exists a neighborhood of the level set $M_c$ in which there is a set of canonical action-angle coordinates 
$\zeta=(\phi_1,\ldots,\phi_n,I_1,\ldots,I_n)$ in which the map takes the form $\mathcal{M}({\phi},{I})=({\phi}^f,{I}^f)$, where:
\begin{equation}
{I}^f={I},\quad\quad {\phi}^f={\phi}+2\pi\nu({I})\quad\operatorname{mod}2\pi, \label{tunemap}
\end{equation}
and the invariants $f_k$ are functions of the action coordinates only, so that:
\begin{equation}
D_\phi\mathcal{F}=0,\quad\quad D\mathcal{F}=\begin{pmatrix}
0 & D_I\mathcal{F}\end{pmatrix}.\label{actionsimp}
\end{equation}
Since we have assumed that $D\mathcal{F}$ is of full rank, it follows from (\ref{actionsimp}) that $D_I\mathcal{F}$ is invertible, and we may apply the result (\ref{TS}) to obtain:
\begin{equation}
{S}=\int_{\gamma}(D_I\mathcal{F})^{-T}d{\phi}. \label{Srefintegral}
\end{equation}
Since the invariants are functions of the action coordinates only, the matrix $D_I\mathcal{F}$ is constant along the integration path $\gamma$, and we need only evaluate an integral of the form:
\begin{equation}
\int_{\gamma}d\phi=\Delta\phi+2\pi m, \label{angleint}
\end{equation}
where $\Delta\phi=(\Delta\phi_1,\ldots,\Delta\phi_n)$ denotes the net change in the angle coordinates $(\phi_1,\ldots,\phi_n)$, when taken to lie in the range $[0,2\pi)$, and $m=(m_1,\ldots,m_n)\in \mathbb{Z}^n$ denotes the number of times the path $\gamma$ winds around the torus with respect to the angles $\phi_1,\ldots,\phi_n$, respectively.  Using (\ref{tunemap}) and (\ref{angleint}) in (\ref{Srefintegral}) gives:
\begin{equation}
{S}=2\pi(D_I\mathcal{F})^{-T}(\nu+m),\quad m\in \mathbb{Z}^n.
\end{equation}
Similarly, we have
\begin{equation}
R_{j k}=\left(\oint_{\gamma_k}(D_I\mathcal{F})^{-T}d{\phi}\right)_j. \label{Rrefintegral}
\end{equation}
By definition, the closed paths $\gamma_k$ $(k=1,\ldots,n)$ form a basis for the group of 1-cycles on $M_c$.  Consider the coordinate curves $\tilde{\gamma}_k:[0,1]\rightarrow M_c$, given in action-angle coordinates by:
\begin{equation}
\tilde{\gamma}_k(t)=(0,\ldots,0,2\pi t,0,\ldots,0), \label{coordcurves}
\end{equation}
where the nontrivial entry corresponds to the $k$th angle coordinate.  Then the paths $\tilde{\gamma}_k$ $(k=1,\ldots,n)$ also form a basis for the group of 1-cycles on $M_c$.  The change of basis is represented by some unimodular integer matrix $U$, so that:
\begin{equation}
\int_{\gamma_k}d\phi=\sum_{l=1}^nU_{kl}\oint_{\tilde{\gamma}_l}d\phi. \label{winding}
\end{equation}
However,
\begin{equation}
\oint_{\tilde{\gamma}_l}d\phi=\int_0^1(2\pi e_l)dt=2\pi e_l.
\end{equation}
It follows that the $l$-th component of (\ref{winding}) is given by:
\begin{equation}
\left(\oint_{\gamma_k}d\phi\right)_l=2\pi U_{kl},
\end{equation}
so using (\ref{Rrefintegral}) gives:
\begin{equation}
R=2\pi(D_I\mathcal{F})^{-T}U^T.
\end{equation}
Since $U^T$ is invertible, it follows that the matrix $R$ is invertible and we have:
\begin{equation}
R^{-1}S=U^{-T}(\nu+m)=U'\nu+m', \label{proofresult}
\end{equation}
where $U'=U^{-T}$ is a unimodular integer matrix, and $m'=U^{-T}m$ is an $n$-vector of integers.  It follows that (\ref{proofresult}) yields the vector of tunes $\nu$ appearing in (\ref{tunemap}), up to an equivalence of the form (\ref{unimodular}).  Coordinate-independence then shows that the same is true of the expression in (\ref{invariant}).

More can be said.  If the basis cycles $\gamma_1,\ldots,\gamma_n$ are initially chosen to be homologous to the coordinate curves $\tilde{\gamma}_1,\ldots,\tilde{\gamma}_n$, then $U'=I_{n\times n}$, and (\ref{proofresult}) correctly yields the vector of tunes $\nu$ modulo 1.  Otherwise, by making a change of coordinates of the form (\ref{aa}), one may transform to action-angle coordinates in which the tunes appearing in (\ref{proofresult}) are equal to those in (\ref{tunemap}), modulo 1.  Thus we may assume, without loss of generality, that the initial action-angle coordinates are chosen such that the coordinate curves (\ref{coordcurves}) are homologous to the basis cycles $\gamma_k$ $(k=1,\ldots,n)$.  In this way, the choice of basis cycles fixes the tunes uniquely mod 1.

This proof also demonstrates that the expression (\ref{invariant}) is independent of the choice of the initial point $\zeta$ and the paths $\gamma$, $\gamma_k$.  This occurs because we can transform to coordinates in which the integrand is constant along these paths, and the path dependence of each integral is determined only by the net change in the angular coordinates along each path.  In particular, the result depends only on the homotopy class of the paths $\gamma$ and $\gamma_k$.

\subsection{Changing the set of invariants}
In the previous subsection, we showed that (\ref{invariant}) correctly produces the dynamical tunes of the map $\mathcal{M}$.  The proof uses the fact that (\ref{invariant}) is invariant under a change of coordinates for the domain of $\mathcal{F}$ (the phase space).  In fact, (\ref{invariant}) is also invariant under a change of coordinates for the range of $\mathcal{F}$ (which is $\mathbb{R}^n$).  More precisely, let $f'=(f_1',\ldots,f_n')$ denote a new set of invariants that is related to the previous set of invariants $f=(f_1,\ldots,f_n)$ through a smooth coordinate transformation $\mathcal{A}:\mathbb{R}^n\rightarrow\mathbb{R}^n$, so that 
\begin{equation}
{f}'=\mathcal{A}({f}). \label{invtransf}
\end{equation}  
Let all quantities expressed in these new coordinates be denoted with $'$.  Then by definition we have:
\begin{equation}
\mathcal{F}'=\mathcal{A}\circ\mathcal{F},\quad\quad D\mathcal{F}'=(D\mathcal{A})(D\mathcal{F}). \label{Ftransf}
\end{equation}
Let the quantity $dv$ be defined as in (\ref{vdef}).  The identity (\ref{invid}) in the primed coordinates is:
\begin{equation}
(D\mathcal{F}')^Tdv'=J'd\zeta'.
\end{equation}
Using (\ref{Ftransf}) and noting that $J'=J$ and $d\zeta'=d\zeta$ gives
\begin{equation}
(D\mathcal{F})^T(D\mathcal{A})^Tdv'=Jd\zeta.
\end{equation}
Multiplying both sides by $(D\mathcal{F}^+)^T$ and noting that this is a left inverse of $(D\mathcal{F})^T$ gives:
\begin{equation}
(D\mathcal{A})^Tdv'=(D\mathcal{F}^+)^TJd\zeta=dv.
\end{equation}
Thus, we have:
\begin{equation}
dv'=(D\mathcal{A})^{-T}dv. \label{forms}
\end{equation}
Since the level sets of $\mathcal{F}$ and $\mathcal{F}'$ coincide, we assume that we use the same paths $\gamma$ and $\gamma_k$ to integrate (\ref{forms}) on both sides of the equality.  Note that $D\mathcal{A}$ is evaluated at the point $\mathcal{F}(\zeta)$, so it depends on the invariants only and is therefore constant along the integration path.  It follows that:
\begin{equation}
{S}'=(D\mathcal{A})^{-T}{S},\quad\quad R'=(D\mathcal{A})^{-T}R,
\end{equation}
and therefore
\begin{equation}
{\nu}'=R^{-1}(D\mathcal{A})^T(D\mathcal{A})^{-T}{S}=R^{-1}S={\nu}.
\end{equation}
This shows that the vector of tunes ${\nu}\in\mathbb{R}^n$ does not change under a transformation (\ref{invtransf}) of the invariants.

One may simplify the proof in the previous subsection as follows.  In addition to using action-angle coordinates to evaluate (\ref{invariant}), one may choose to transform the invariants $(f_1,\ldots,f_n)$ to the set of action coordinates $(I_1,\ldots,I_n)$ using an invertible transformation $\mathcal{A}({f})={I}$.  Using these coordinates for the domain and range of $\mathcal{F}$, we have $D_I\mathcal{F}'=I_{n\times n}$, the identity, and the integrals (\ref{Srefintegral},\ref{Rrefintegral}) take a trivial form.  We chose not to take this approach, in order to illustrate explicitly the path independence of the separate factors ${S}$ and $R$.

\section{Frequencies of Hamiltonian flows}
Let $\mathcal{M}$ denote the period-1 map associated with an integrable Hamiltonian $H$.  Expressing the dynamics in action-angle form, we have:
\begin{equation}
I(t)=I(0),\quad {\phi}(t)={\phi}(0)+{\omega}({I(0)})t,
\end{equation}
where the frequency vector $\omega=(\omega_1,\ldots,\omega_n)$ is given by:
\begin{equation}
\omega_k=\frac{\partial H}{\partial I_k}. \label{freqdef}
\end{equation}
The period-1 map is given by $\mathcal{M}({\phi},I)=({\phi}^f,I^f)$, where
\begin{equation}
I^f=I,\quad {\phi}^f={\phi}+2\pi\nu({I}),\quad {\nu}=\frac{{\omega}}{2\pi}.
\end{equation}
It follows that we may apply the result for integrable maps (\ref{invariant}) to extract the frequency vector ${\omega}$ without knowledge of the actions $I$ that appear in (\ref{freqdef}).

Of the many available choices for the path $\gamma$, we may choose an integral curve of the Hamiltonian flow.  Along this curve,
\begin{equation}
\frac{d{\zeta}}{dt}={J}\nabla H({\zeta}).
\end{equation}
Assume that $H$ is given by some function $G$ of the invariants, so that $H=G\circ\mathcal{F}$.  Then
\begin{equation}
DH=(DG)(D\mathcal{F}), \label{Hdef}
\end{equation}
and
\begin{equation}
\frac{d{\zeta}}{dt}=J(D\mathcal{F})^T(DG)^T.
\end{equation}
Using this as the path $\gamma$ in (\ref{invariant}), and noting that application of the map corresponds to moving from $t=0$ to $t=1$:
\begin{equation}
{S}=\int_0^1(D\mathcal{F}^+)^T(D\mathcal{F})^T(DG)^Tdt.
\end{equation}
Since $(D\mathcal{F}^+)^T$ is a left inverse of $(D\mathcal{F})^T$, and the matrix $DG$ is constant along the path, it follows that:
\begin{equation}
{S}=DG^T,\quad\quad {\nu}=R^{-1}DG^T. \label{freq1}
\end{equation}
In the special case that $H=f_1$, then $DG^T={e}_1$ and
\begin{equation}
{\omega}=2\pi R^{-1}{e}_1, \label{freq2}
\end{equation}
where ${e}_1=(1,0,0,\ldots,0)^T$.  Note that the result (\ref{freq1}) no longer requires explicit knowledge of the period-1 map $\mathcal{M}$, which has been eliminated in favor of the Hamiltonian $H$.

Let us check the coordinate-invariant expression (\ref{freq1}) by evaluating the matrix $R$ using action-angle coordinates.  In these coordinates,
\begin{equation}
R_{j k}=\left(\oint_{\gamma_k}(D_I\mathcal{F})^{-T}d{\phi}\right)_j.
\end{equation}
Since the matrix $D_I\mathcal{F}$ is constant along the integration path, it follows that:
\begin{equation}
R=2\pi(D_I\mathcal{F})^{-T}.
\end{equation}
But then:
\begin{equation}
{\nu}=R^{-1}{S}=\frac{1}{2\pi}(D_I\mathcal{F})^T(DG)^T. \label{HamTest}
\end{equation}
Finally, evaluating expression (\ref{Hdef}) in terms of its $n\times n$ blocks gives:
\begin{equation}
\begin{pmatrix}
D_{\phi}H & D_{I}H\end{pmatrix}=
DG\begin{pmatrix} D_{\phi}\mathcal{F} & D_I\mathcal{F}\end{pmatrix},
\end{equation}
so that:
\begin{equation}
D_IH=(DG)(D_I\mathcal{F}).
\end{equation}
Using this result in (\ref{HamTest}) and multiplying by $2\pi$ then gives:
\begin{equation}
{\omega}=(D_IH)^T,\quad\text{or}\quad \omega_j=\frac{\partial H}{\partial I_j},
\end{equation}
which is (\ref{freqdef}).

\section{Numerical Examples}
To illustrate the application of (\ref{invariant}) using practical examples, the results of this paper were used to determine: 1) the dynamical frequencies of one nonlinear Hamiltonian flow, and 2) the tunes of one nonlinear symplectic map, both defined on the phase space $\mathbb{R}^4$.  Appendix C illustrates in detail how (\ref{invariant}) can also be used to correctly produce the tunes of a stable linear symplectic map of arbitrary dimension.
\subsection{Integrable H\'enon-Heiles Hamiltonian}
Consider the Hamiltonian given by (for $\lambda>0$):
\begin{equation}
H=\frac{1}{2}\left(p_x^2+p_y^2+x^2+y^2\right)+\lambda\left(x^2y+\frac{y^3}{3}\right). \label{HHint}
\end{equation}
This is the usual H\'enon-Heiles Hamiltonian \cite{Henon}, except that the sign of the $y^3$ term is reversed.  It is known that (\ref{HHint}) is
integrable, with two invariants of the form \cite{HenonInt1,HenonInt2}:
\begin{equation}
f_1=H,\quad\quad f_2=p_xp_y+xy+\lambda\left(xy^2+\frac{x^3}{3}\right). \label{HHinv}
\end{equation}
An analysis of (\ref{HHinv}) shows that an invariant level set $M_c$ for some $c\in\mathbb{R}^2$ contains a connected component $M_c^0$ near the origin that is regular and compact provided that:
\begin{align}
0\leq c_1-c_2\leq \frac{1}{6\lambda^2}, \quad 0\leq c_1+c_2\leq\frac{1}{6\lambda^2}.
\end{align}
For orbits on $M_c^0$, we wish to evaluate the characteristic frequency vector ${\omega}=(\omega_1,\omega_2)^T$ using (\ref{freq2}).  

To evaluate the path integrals appearing in the matrix $R$, we need to choose two basis cycles $\gamma_1$, $\gamma_2$ lying in the two-dimensional surface $M_c^0$.  One approach is to consider the curve obtained by intersecting $M_c^0$ with the hyperplane $y=kx$ $(k\in\mathbb{R})$.  Using (\ref{HHinv}) to solve for $p_x$ and $p_y$ locally in terms of the coordinates $x$, $y$ and setting $y=kx$ gives the parameterized curve segment:
\begin{equation}
t\mapsto (t,kt,p_x(t),p_y(t)),
\end{equation}
where $p_x(t)$ is given by:
\begin{align}
p_x(t)&=\pm\sqrt{\frac{1}{2}(c_1+c_2)-\frac{1}{4}(k+1)^2t^2-\frac{\lambda}{6}(k+1)^3t^3} \notag \\
&\pm\sqrt{\frac{1}{2}(c_1-c_2)-\frac{1}{4}(k-1)^2t^2-\frac{\lambda}{6}(k-1)^3t^3}, \label{pxsolve}
\end{align}
and the signs of the two terms may be chosen independently.  In each case, $p_y(t)$ is given by reversing the sign of the second term in (\ref{pxsolve}).
To construct the cycle $\gamma_1$, one must then paste together curve segments that utilize the appropriate signs in (\ref{pxsolve}) to produce a closed path.  For convenience, the closed path $\gamma_2$ is obtained using the same procedure, for the choice of intersecting hyperplane $y=-kx$.  Independence of the two cycles $\gamma_1$ and $\gamma_2$ will be explored momentarily.

\begin{figure}
\begin{center}
\includegraphics[width=2.7in]{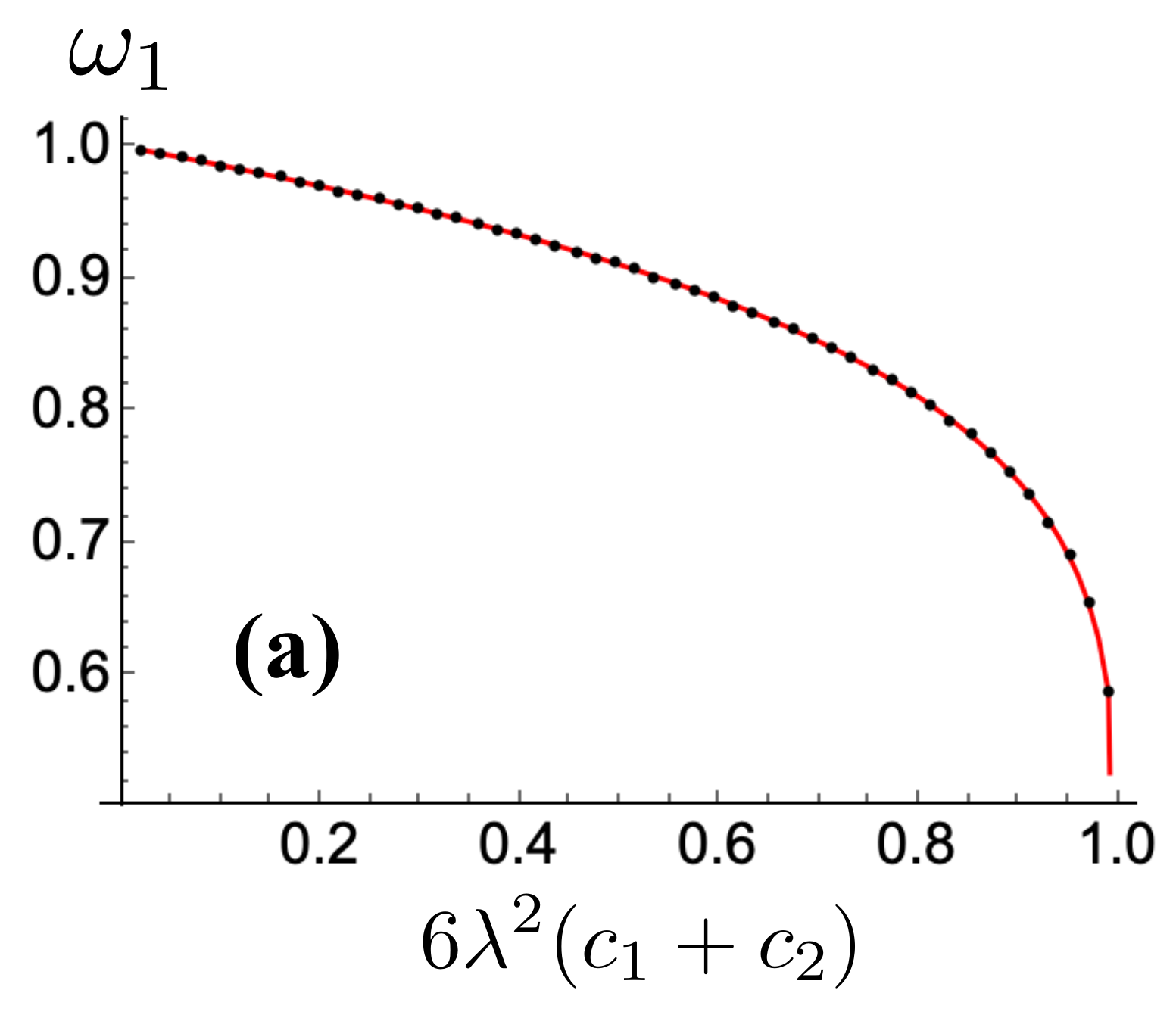}
\includegraphics[width=2.7in]{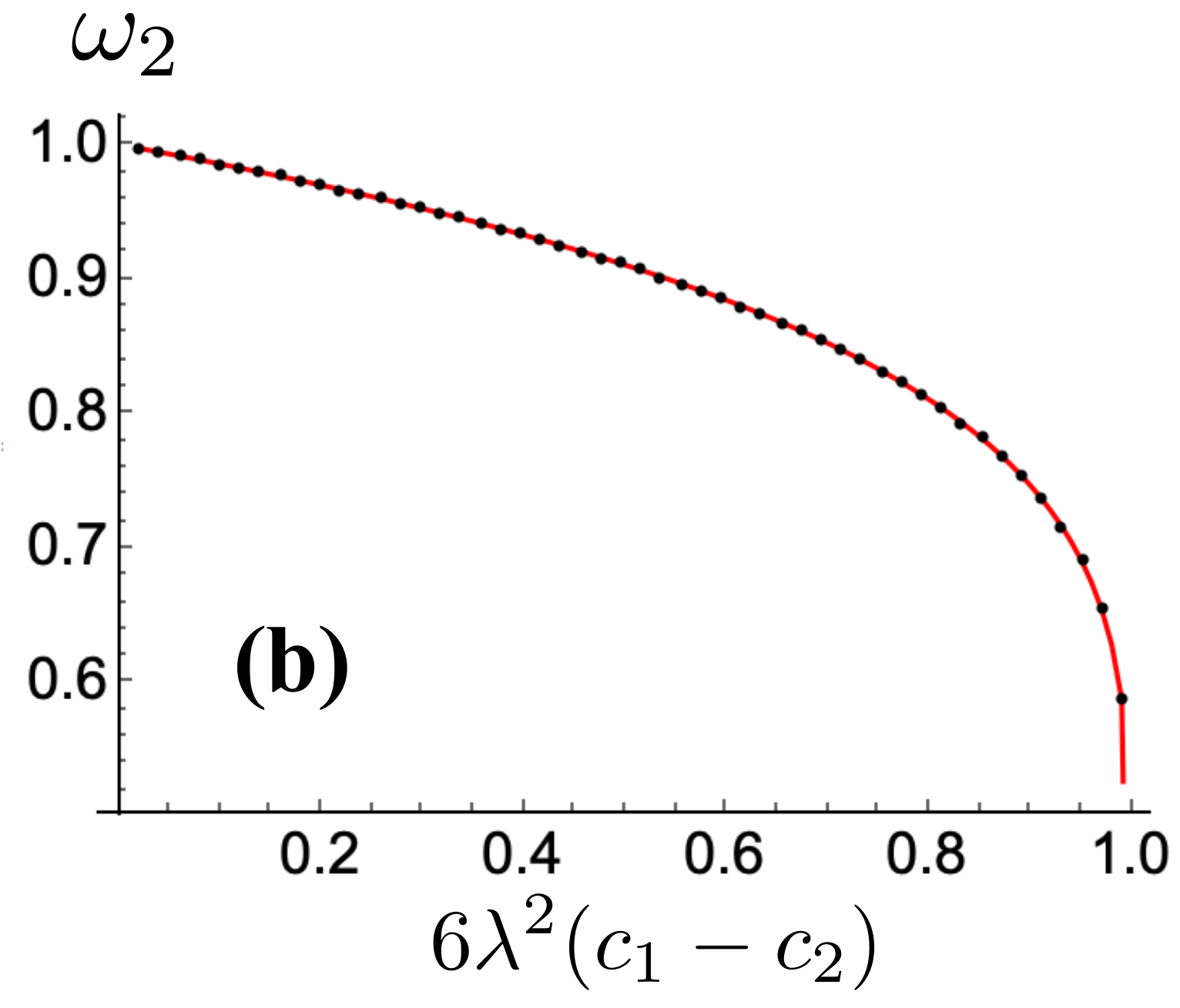}
\end{center}
\caption{Frequencies of the Hamiltonian (\ref{HHint}) with $\lambda=1$, shown for the level set $M_c^0$ defined by $(f_1,f_2)=(c_1,c_2)$.  Dots correspond to the analytical expression given in (\ref{freqexact}), while solid curves correspond to the result obtained using (\ref{invariant}).  (a) The value $\omega_1$ is shown for $6\lambda^2(c_1-c_2)=1/2$.  (b) The value $\omega_2$ is shown for $6\lambda^2(c_1+c_2)=1/2$.  In both cases, a separatrix is approached as the horizontal axis approaches 1.}
\label{HenonResult}
\end{figure}

 In the coordinates $(x,y,p_x,p_y)$, note that the Jacobian matrix of the momentum mapping is given by:
\begin{equation}
D\mathcal{F}=\begin{pmatrix}
x+2\lambda xy & y+\lambda(x^2+y^2) & p_x & p_y \\
y+\lambda(x^2+y^2) & x+2xy\lambda & p_y & p_x
\end{pmatrix},
\end{equation}
and its Moore-Penrose inverse (\ref{MooreP}) can be evaluated explicitly.  Alternatively, we may use only the $2\times2$ momentum block $D_p\mathcal{F}$ by applying (\ref{TS}), provided we avoid points where $p_x=0$ or $p_y=0$.  Evaluating the integrals in (\ref{invariant}) numerically along the paths $\gamma_1$ and $\gamma_2$ to produce the matrix $R$, and using (\ref{freq2}) to produce the frequency vector $\omega$ yields the results shown in Fig. \ref{HenonResult}

This system can also be solved exactly.
Note that by making the symplectic coordinate transformation:
\begin{align}
q_1&=\frac{1}{\sqrt{2}}(y+x),\quad p_1=\frac{1}{\sqrt{2}}(p_y+p_x), \\
q_2&=\frac{1}{\sqrt{2}}(y-x),\quad p_2=\frac{1}{\sqrt{2}}(p_y-p_x),
\end{align}
the Hamiltonian decouples as:
\begin{equation}
H=H_1+H_2,\quad\quad H_j=\frac{1}{2}\left(p_j^2+q_j^2\right)+\frac{\lambda\sqrt{2}}{3} q_j^3,
\end{equation}
and the invariants take the form:
\begin{equation}
f_1=H_1+H_2,\quad\quad f_2=H_1-H_2.
\end{equation}
Periodic motion in the coordinate $q_j$ $(j=1,2)$ occurs between two turning points $q_{j}^{min}$, $q_{j}^{max}$ when: 
\begin{equation}
0\leq H_j\leq \frac{1}{12\lambda^2}=H_{max},
\end{equation}
with period given by:
\begin{equation}
T_j=\oint \left(\frac{dq_j}{dt}\right)^{-1}dq_j = 2\int_{{q_j^{min}}}^{q_j^{max}}\frac{dq_j}{\sqrt{2H_j-q_j^2-2\lambda\sqrt{2} q_j^3/3}}. \notag
\end{equation}
The corresponding frequency $\omega_j=2\pi/T_j$ is given explicitly by:
\begin{equation}
\omega_j=\frac{\pi\sqrt{\zeta_{bj}-\zeta_{aj}}}{\sqrt{6}K\left(\frac{\zeta_{cj}-\zeta_{bj}}{\zeta_{aj}-\zeta_{bj}}\right)}, \label{freqexact}
\end{equation}
where $K$ denotes the complete elliptic integral of the first kind, and $\zeta_{aj}$, $\zeta_{bj}$, $\zeta_{cj}$ denote the three roots of the polynomial:
\begin{equation}
P_j(\zeta)=2\zeta^3+3\zeta^2-(H_j/H_{max}),
\end{equation}
ordered such that $\zeta_{aj}<\zeta_{bj}<0<\zeta_{cj}$ for $j=1,2$.

\begin{figure}
\begin{center}
\includegraphics[width=2.7in]{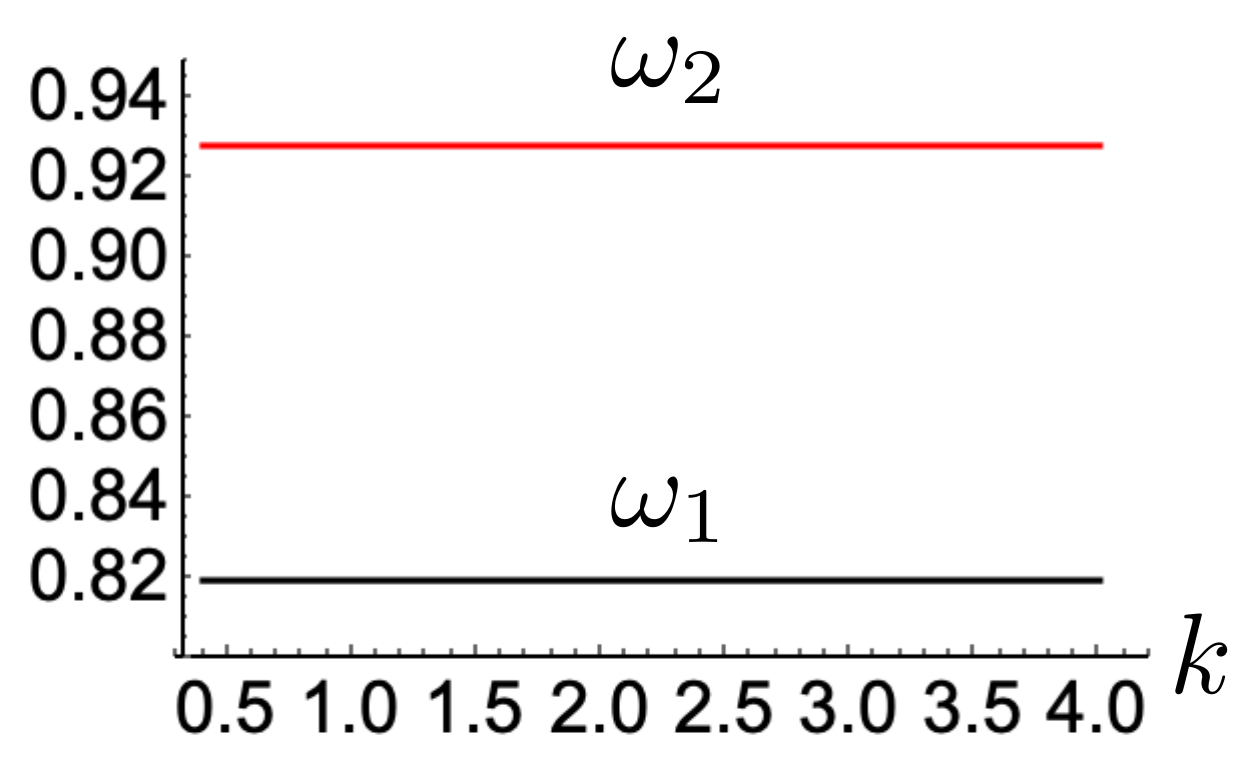}
\end{center}
\caption{Demonstration that the frequencies of the Hamiltonian (\ref{HHint}) ($\lambda=1$) obtained using (\ref{freq2}) are unchanged under deformation of the cycles $\gamma_1$ and $\gamma_2$.  These are defined by intersection of the level set $M_c^0$ with the hyperplanes $y=kx$ and $y=-kx$, respectively.  The results are shown for the case $c_1=0.1$, $c_2=0.03$.}
\label{HenonInv}
\end{figure}

Figure \ref{HenonResult} shows a comparison between the result obtained by numerically evaluating the path integrals in (\ref{freq2}) and the exact solution in (\ref{freqexact}).  This result is shown for $k=1/2$.  By varying $k$, one may study the dependence on the choice of cycles $\gamma_1$ and $\gamma_2$.  For example, Fig. \ref{HenonInv} shows that the frequencies $\omega_1$, $\omega_2$ on the level set $(c_1,c_2)=(0.1,0.03)$ are independent of $k$, for $0.4<k<4.5$.  Beyond this range, the two cycles obtained by intersecting $M_c^0$ with the hyperplanes $y=kx$ and $y=-kx$ fail to be independent, and the matrix $R$ is not invertible.  In this case, at least one of the two cycles must be modified if (\ref{freq2}) is to be used.


\subsection{Integrable 4D McMillan Map}
Consider the symplectic map $\mathcal{M}:\mathbb{R}^4\rightarrow \mathbb{R}^4$ given by $\mathcal{M}(x,y,p_x,p_y)=(x^f,y^f,p_x^f,p_y^f)$, where:
\begin{subequations}\label{McMilMap}
\begin{align}
x^f&=p_x,\quad\quad p_x^f=-x+\frac{ap_x}{1+b(p_x^2+p_y^2)}, \\
y^f&=p_y,\quad\quad p_y^f=-y+\frac{ap_y}{1+b(p_x^2+p_y^2)},
\end{align}
\end{subequations}
and $a,b>0$.  This is a 4D analogue of the so-called McMillan mapping \cite{McMillan}.  It is known that (\ref{McMilMap}) is integrable, with two invariants of the form:
\begin{subequations}\label{McMilInv}
\begin{align}
f_1&=x^2+y^2+p_x^2+p_y^2-a(xp_x+yp_y)  \\
&\quad\quad\quad\quad\quad +b(xp_x+yp_y)^2, \notag \\
f_2&=xp_y-yp_x.
\end{align}
\end{subequations}
We wish to evaluate the tunes of this map using (\ref{invariant}).

\begin{figure}
\begin{center}
\includegraphics[width=2.7in]{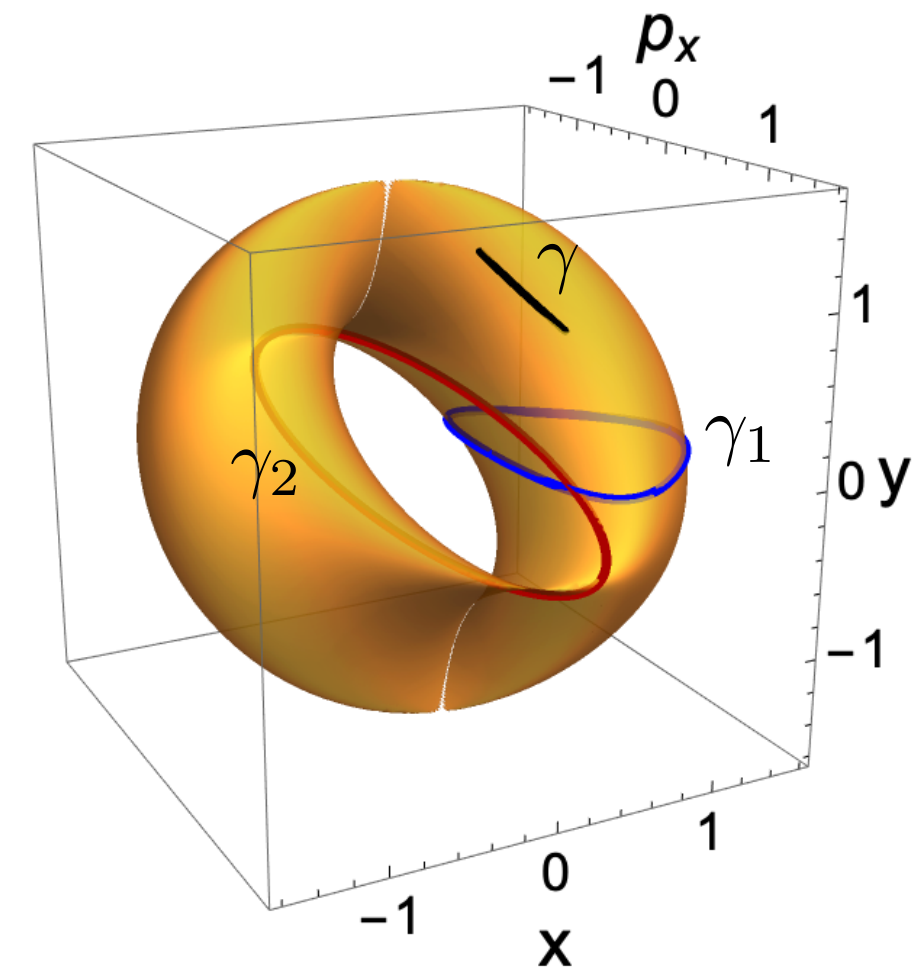}
\end{center}
\caption{(Orange) Level set $(f_1,f_2)=(2,0.5)$ of the 4D McMillan map (\ref{McMilMap}) with $a=1.6$, $b=1$.  The apparent self-intersections of the 2D surface are an artifact of projection into $\mathbb{R}^3$.  This is shown together with examples of basis cycles $\gamma_1$ and $\gamma_2$ and the path $\gamma$ used to evaluate the tunes $\nu_1$, $\nu_2$ from (\ref{invariant}).}
\label{McMilCycles}
\end{figure}

The cycles $\gamma_1$ and $\gamma_2$ can be defined, as before, by taking the intersection of $M_c$ with hypersurfaces of the form $G_j(x,y)=0$ for smooth functions $G_j$ $(j=1,2)$, chosen to make $\gamma_1$ and $\gamma_2$ independent.  One must also choose an arbitrary initial point $\zeta\in M_c$ and a path $\gamma$ to its image $\mathcal{M}(\zeta)$.
An example of a regular invariant level set is shown in Fig. \ref{McMilCycles}, together with two independent basis cycles $\gamma_1$ and $\gamma_2$, and the path $\gamma$.

In the coordinates $(x,y,p_x,p_y)$, note that the Jacobian matrix of the momentum mapping is given by:
\begin{equation}
D_q\mathcal{F}=\begin{pmatrix}
-ap_x+2(x+bp_x\tau) & -ap_y+2(y+bp_y\tau)  \\
p_y & -p_x
\end{pmatrix}, \notag
\end{equation}
\begin{equation}
D_p\mathcal{F}=\begin{pmatrix}
-ax+2(p_x+bx\tau) & -ay+2(p_y+by\tau)  \\
-y & x
\end{pmatrix}, \notag
\end{equation}
where $\tau=xp_x+yp_y$.  Using these results, the integrals in (\ref{invariant}) can be evaluated numerically to obtain the rotation vector $\nu$ as a function of the two invariants.

\begin{figure}
\begin{center}
\includegraphics[width=2.7in]{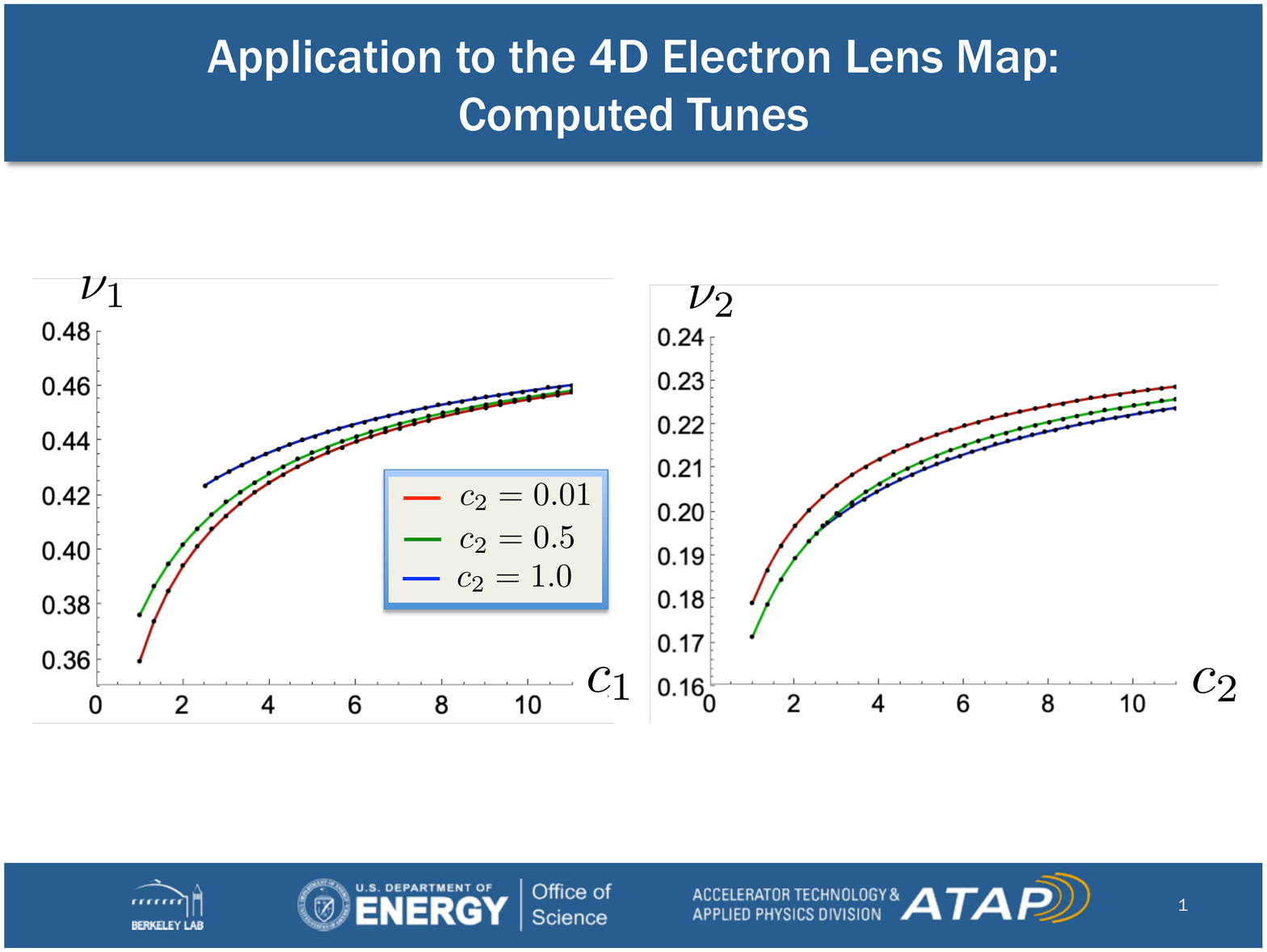}
\includegraphics[width=2.7in]{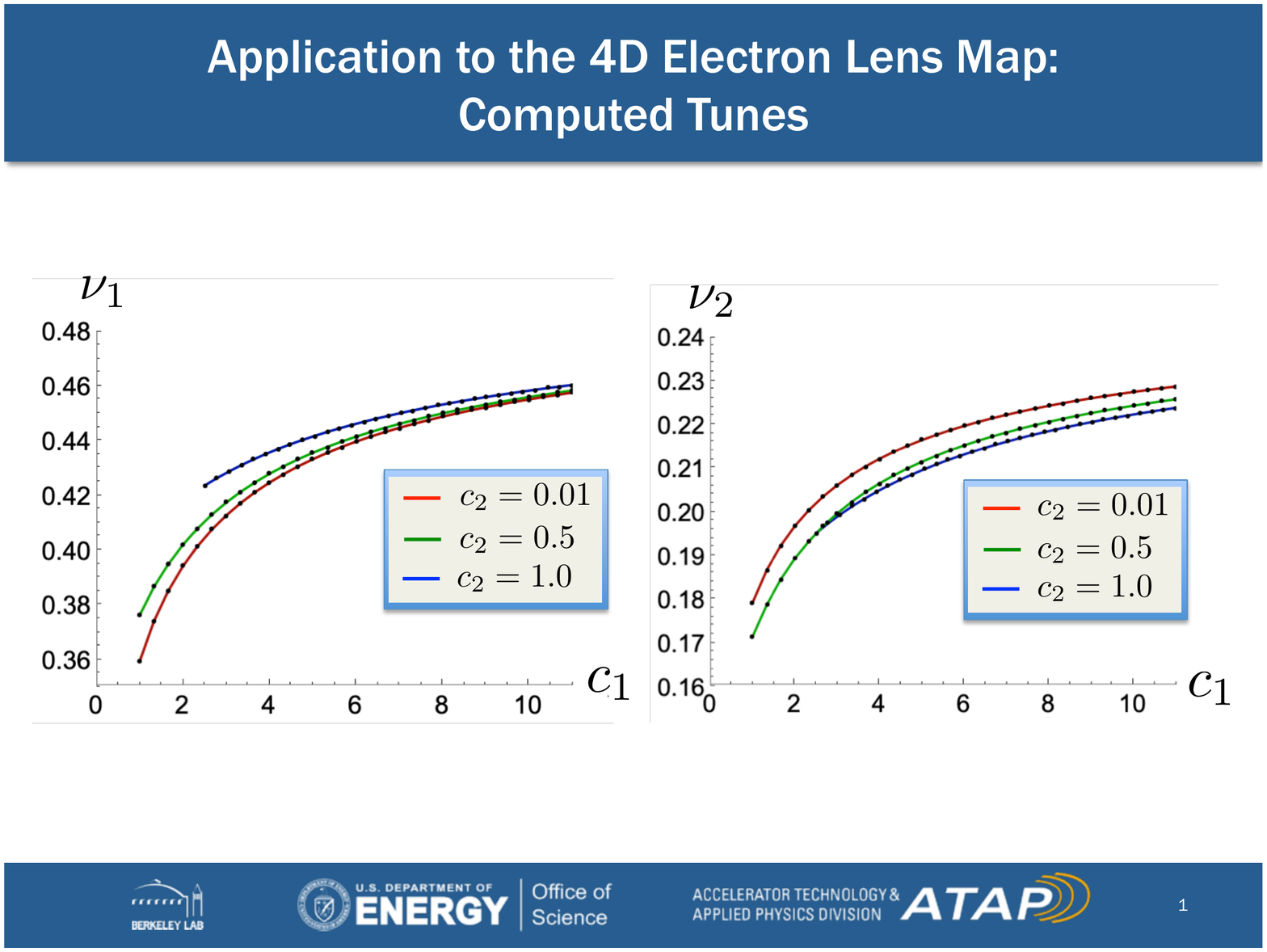}
\end{center}
\caption{Tunes $\nu_1$, $\nu_2$ of the 4D McMillan map (\ref{McMilMap}) with $a=1.6$, $b=1$, shown for the invariant level set defined by $(f_1,f_2)=(c_1,c_2)$.  Dots correspond to the analytical expression given in \cite{Zolkin}, while solid curves correspond to the result obtained using (\ref{invariant}).  Compare Figure 5 of \cite{Zolkin}.}
\label{McMilCf}
\end{figure}

This system can also be solved exactly \cite{Zolkin}.  Figure \ref{McMilCf} shows a comparison between the exact solution provided in \cite{Zolkin} and the solution obtained using the above procedure.  The agreement confirms that the tunes can be accurately determined from (\ref{invariant}), without the construction of action-angle coordinates or knowledge of a coordinate system in which the dynamics is separable.

\section{Conclusions}
Integrable Hamiltonian systems and symplectic maps play important roles in many areas of science, as well as providing an active area of contemporary mathematical research \cite{Bolsinov}.  However, the standard techniques for exact solution of these systems are difficult to apply, except in the simplest cases.
This paper provides an explicit formula (\ref{invariant}) that connects the $n$ tunes of an integrable symplectic map (on a phase space of dimension 2$n$) with its $n$ invariants of motion.  The same formula can be used to extract the $n$ dynamical frequencies of a Hamiltonian flow (Section VI).  By construction, the formula is invariant under a canonical (symplectic) change of coordinates and can be expressed in a geometric form that is coordinate-free.   The construction of action-angle coordinates is not required. 

This formula is consistent with an expression previously obtained for 2D integrable symplectic maps \cite{Zolkin}, and it reproduces exactly known results for dynamical frequencies that have been independently obtained for several nonlinear benchmark problems (Section VII).  A demonstration that this result correctly reproduces the tunes of a linear symplectic map of any dimension is found in Appendix C, and additional special cases of low dimension are treated in Appendix D.

In practice, this formula can be used to extract the dynamical frequencies of the orbits of an integrable system without the need for numerical tracking, which is especially useful when studying the dependence of the dynamical frequencies on the choice of the initial condition or system parameters.  Evaluation of (\ref{invariant}) requires only that one parameterize a set of paths in the invariant level set, which is often done by solving locally for one of the phase space variables in terms of the others.  Note that this result can also be applied to extract approximate dynamical frequencies of orbits (of a symplectic map or a Hamiltonian flow) when a sufficient number of approximate invariants are known.

Most importantly, the expression (\ref{invariant}) captures, in a precise way, the connection between the geometry of an integrable system and its dynamical behavior, providing first-principles insight into the physics of such systems.

\section{Acknowledgments}
The authors thank A. Valishev and the IOTA collaboration team at Fermilab for discussions.
This work was supported by the Director, Office of Science of the U.S. Department of Energy under Contracts No. DE-AC02-05CH11231 and DE-AC02-07CH11359, and made use of computer resources at the National Energy Research Scientific Computing Center.  The authors acknowledge support from the U.S. DOE Early Career Research Program under the Office of High Energy Physics.  

\section*{Appendix A:  Cycles on the Torus}
The closed paths $\gamma_k$ $(k=1,\ldots,n)$ appearing in (\ref{invariant}) must lie within the invariant level set $M_c$, and they must form a basis for the group of 1-cycles on $M_c$.  A proper discussion of the latter condition requires the use of (singular) homology \cite{Hatcher}.  However, intuition for this condition can be obtained by visualizing several examples for the special case when $n=2$ (dimension 4).

In this case, each regular level set $M_c$ can be smoothly deformed into the standard 2-torus, defined by:
\begin{equation}
\mathbb{T}^2=\{(q_1,q_2,p_1,p_2)\in\mathbb{R}^{4}:(\forall j)q_j^2+p_j^2=1 \}. \notag
\end{equation}
Let $q:\mathbb{R}^2\rightarrow\mathbb{T}^2$ denote the function given by:
\begin{equation}
q(t_1,t_2)=(\cos 2\pi t_1,\cos 2\pi t_2,\sin 2\pi t_1,\sin 2\pi t_2).
\end{equation}
Let $\gamma:[a,b]\rightarrow\mathbb{T}^2$ be any continuous path with $\gamma(a)=\gamma(b)$.  A {\it lift}
of $\gamma$ is a continuous map $\tilde{\gamma}:[a,b]\rightarrow\mathbb{R}^2$ such that $\gamma=q\circ\tilde{\gamma}$.
For any closed path $\gamma$, define its {\it index} by:
\begin{equation}
[\gamma]=\tilde{\gamma}(b)-\tilde{\gamma}(a)\in \mathbb{Z}^2.
\end{equation}
It can be verified that the index does not depend on the specific choice of the lift $\tilde{\gamma}$.  It is also invariant under continuous deformations of the path $\gamma$.
Intuitively, $[\gamma]$ is a pair of integers denoting how many times the path $\gamma$ ``winds around" the torus with respect to each of its two ``holes". 
 Two closed paths $\gamma_1$ and $\gamma_2$ will be said to form a {\it basis}
for the group of 1-cycles on $\mathbb{T}^2$ when $[\gamma_1]$ and $[\gamma_2]$ form a basis for the additive group $\mathbb{Z}^2$ over the integers.  

\begin{figure}
\begin{center}
\includegraphics[width=1.6in]{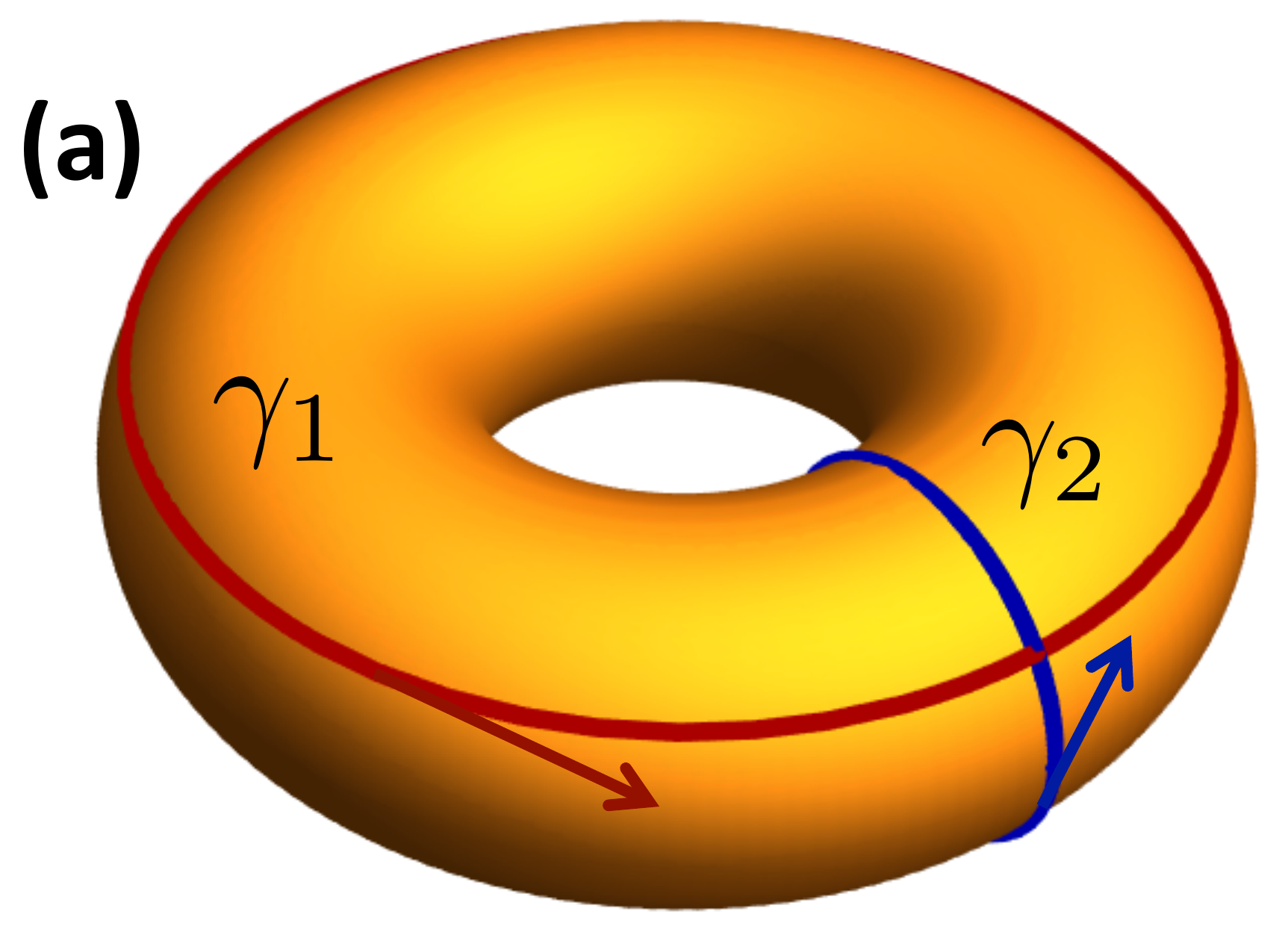}
\includegraphics[width=1.6in]{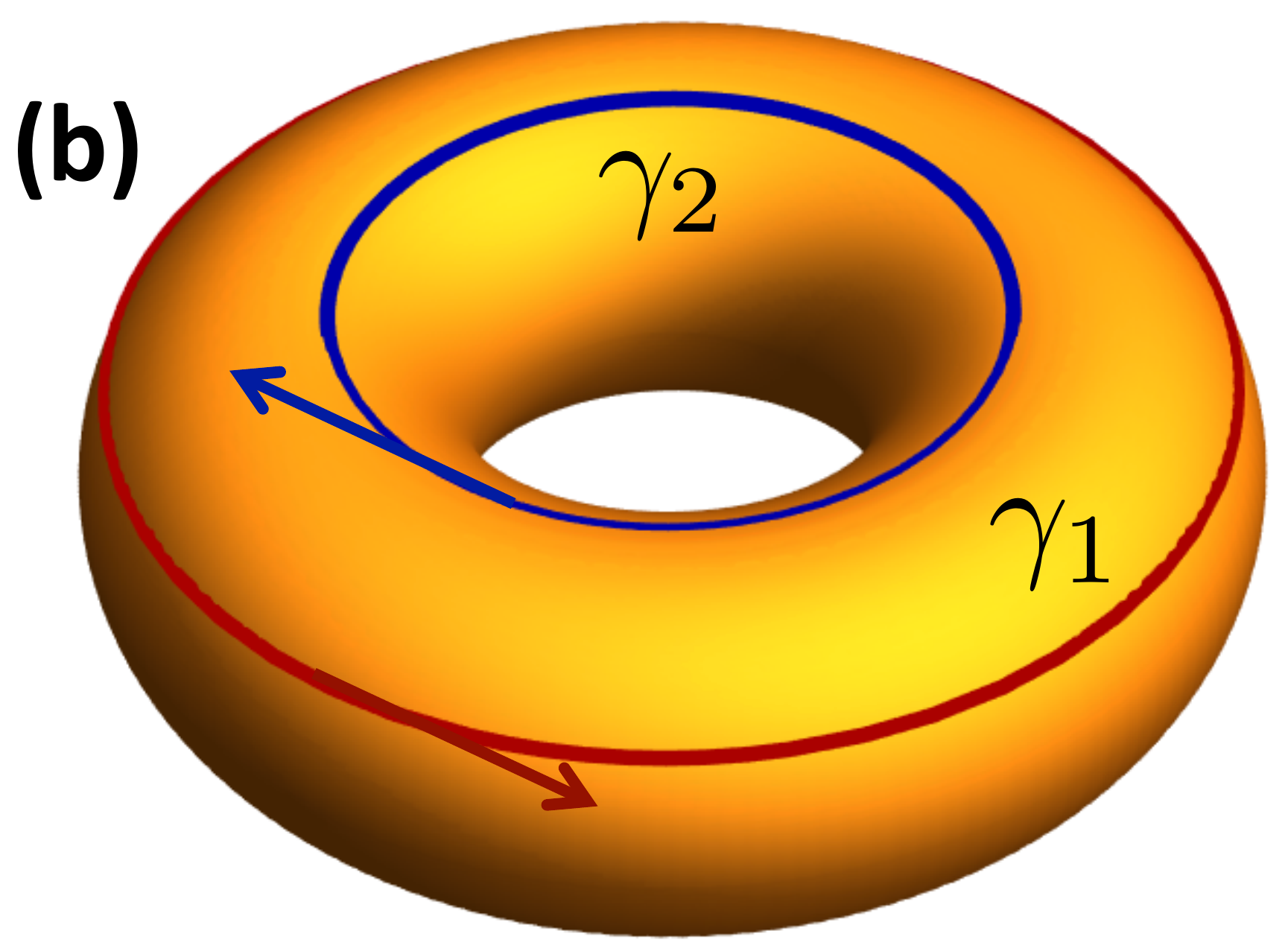}
\includegraphics[width=1.6in]{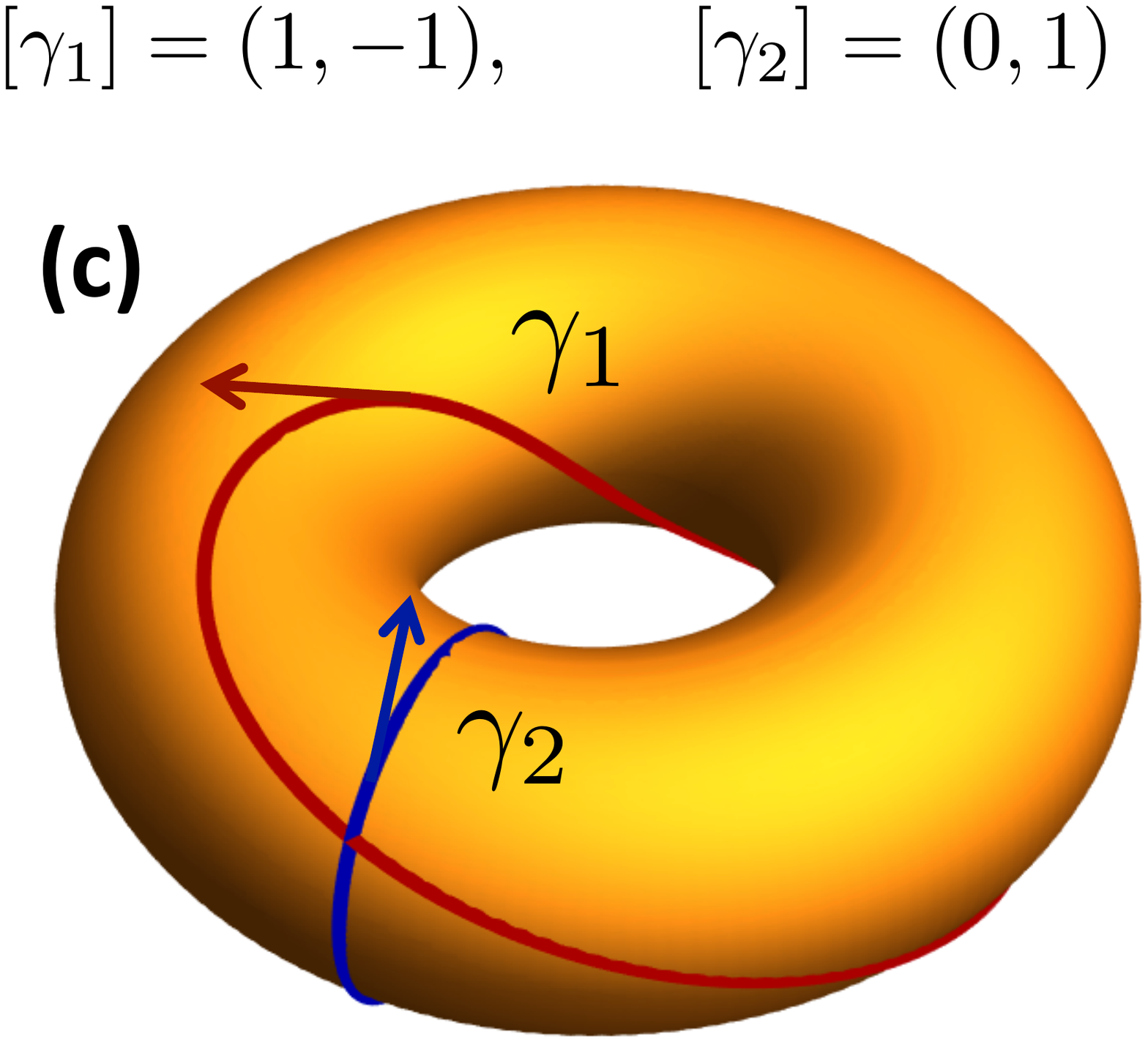}
\end{center}
\caption{Examples of 1-cycles on the torus $\mathbb{T}^2$.  One of the two holes has been made larger than the other, in order to embed the torus in $\mathbb{R}^3$ without self-intersection.  (a) Two basis cycles with $[\gamma_1]=(1,0)$ and $[\gamma_2]=(0,1)$.  (b) Two cycles that do not form a basis, with $[\gamma_1]=(1,0)$, $[\gamma_2]=(-1,0)$.  (c) Two basis cycles with $[\gamma_1]=(1,-1)$ and $[\gamma_2]=(0,1)$.}
\label{Torus}
\end{figure}

The simplest example of a basis on $\mathbb{T}^2$ is shown in Fig. \ref{Torus}(a).  The paths $\gamma_1$ and $\gamma_2$ can be represented by the lifts:
\begin{align}
\tilde{\gamma}_1(t)=(t,0),\quad\tilde{\gamma}_2=(0,t),\quad 0\leq t\leq 1,
\end{align}
so that $[\gamma_1]=(1,0)$ and $[\gamma_2]=(0,1)$.  
Any two paths that can be obtained by continuous deformation of the paths $\gamma_1$ and $\gamma_2$ also results in a basis.

Fig. \ref{Torus}(b) illustrates an example of two closed paths that do not form a basis on $\mathbb{T}^2$.  In fact, if $-\gamma_2$ denotes the path $\gamma_2$ transversed in the opposite direction,
then the path $-\gamma_2$ can be continuously deformed into $\gamma_1$.

A less obvious example of a basis on $\mathbb{T}^2$ is given in Fig. \ref{Torus}(c).  In this example, $[\gamma_1]=(0,1)$ and $[\gamma_2]=(1,-1)$.  The number of such possible bases is infinite, but
bases whose cycles have larger index become increasingly difficult to visualize.

\section*{\label{sec:AppB} Appendix B:  Properties of the Moore-Penrose inverse}
The Poisson bracket condition that $\{f_j,f_k\}=0$ for all $j$, $k$ is equivalent to the matrix identity:
\begin{equation}
(D\mathcal{F})J(D\mathcal{F})^T=0. \label{PB}
\end{equation}
It follows from (\ref{MooreP}) and (\ref{PB}) that $D\mathcal{F}^+$ satisfies the two conditions:
\begin{equation}
(D\mathcal{F})(D\mathcal{F}^+)=I_{n\times n},\quad (D\mathcal{F}){J}(D\mathcal{F}^+)=0. \label{MPIcond}
\end{equation}
%In fact, these conditions determine the matrix $D\mathcal{F}^+$ uniquely.
Consider the linear map corresponding to $(D\mathcal{F}^+)(D\mathcal{F})$.  This map is a linear projection since:
\begin{equation}
(D\mathcal{F}^+D\mathcal{F})^2=D\mathcal{F}^+D\mathcal{F}.
\end{equation}
  We examine its null space ($\operatorname{ker}$) and range ($\operatorname{im}$).  Using the leftmost identity in (\ref{MPIcond}), we obtain:
\begin{equation}
\operatorname{ker}(D\mathcal{F}^+D\mathcal{F})=\operatorname{ker}(D\mathcal{F}). \label{ker1}
\end{equation}
Similarly, it follows from the rightmost identity in (\ref{MPIcond}) that:
\begin{equation}
\operatorname{im}(D\mathcal{F}^+D\mathcal{F})\subseteq \operatorname{ker}(D\mathcal{F}{J}). \label{subspace}
\end{equation}
It is straightforward to verify that
\begin{equation}
\operatorname{ker}(D\mathcal{F}{J})=J\operatorname{ker}(D\mathcal{F}) \label{ker2}
\end{equation}
and since $J$ is invertible,
\begin{equation}
\operatorname{dim}(J\operatorname{ker}(D\mathcal{F}))=\operatorname{dim}(\operatorname{ker}(D\mathcal{F})). \label{ker3}
\end{equation}
Since $\operatorname{rank}(D\mathcal{F})=n$ by assumption, it follows by the rank-nullity theorem that
$\operatorname{dim}(\operatorname{ker}(D\mathcal{F}))=n$.
By (\ref{ker1}-\ref{ker3}), the two subspaces in (\ref{subspace}) have the same dimension $n$, and it follows that they coincide:
\begin{equation}
\operatorname{im}(D\mathcal{F}^+D\mathcal{F})=J\operatorname{ker}(D\mathcal{F}).
\end{equation}
Thus, at every point in the phase space $M$ we have the direct-sum decomposition:
\begin{equation}
\mathbb{R}^{2n}=\operatorname{ker}(D\mathcal{F})\oplus J\operatorname{ker}(D\mathcal{F}), \label{splitting}
\end{equation}
and the projection $P$ onto the second summand is given by:
\begin{equation}
P=(D\mathcal{F}^+)(D\mathcal{F}). \label{project}
\end{equation}
The two conditions (\ref{MPIcond}) therefore determine $D\mathcal{F}^+$ uniquely.  For if $B$ is any matrix satisfying the two conditions (\ref{MPIcond}), then for any vector $\zeta\in\mathbb{R}^{2n}$,
\begin{equation}
(D\mathcal{F})JB\zeta=0,
\end{equation}
so that $B\zeta$ lies in $\operatorname{ker}(D\mathcal{F}J)=J\operatorname{ker}(D\mathcal{F})$, and therefore:
\begin{equation}
B\zeta=PB\zeta=(D\mathcal{F}^+)(D\mathcal{F})B\zeta=(D\mathcal{F}^+)\zeta.
\end{equation}
The results (\ref{splitting}-\ref{project}) are used in Section \ref{sec:path}.

\section*{Appendix C:  Treatment of Linear Maps}
Consider a linear symplectic map on the phase space $M=\mathbb{R}^{2n}$, represented by a $2n\times 2n$ real symplectic matrix $R$.  Suppose that the $2n$ eigenvalues of $R$ are distinct and lie on the unit circle.  It follows that the eigenvalues of $R$ occur in complex-conjugate pairs, and one may select $n$ eigenvalues $\lambda_j$ and (complex) eigenvectors $\psi_j$ so that for $j=1,\ldots,n$: 
\begin{equation}
R\psi_j=\lambda_j\psi_j,\quad\quad R\bar{\psi}_j=\bar{\lambda}_j\bar{\psi}_j,\quad\quad |\lambda_j|=1. \label{Rconditions}
\end{equation}
Following \cite{Dragt}, we introduce the angular bracket notation:
\begin{equation}
\langle{u,v\rangle}=-i\bar{u}^TJv,\quad \quad u,v\in\mathbb{C}^{2n}.
\end{equation}
Then the eigenvectors $\psi_j$ may be indexed and normalized such that for $l,m=1,\ldots,n$:
\begin{subequations}\label{orthonormal}
\begin{align}
\langle{\psi_l,\psi_m\rangle}&=\delta_{l,m}, \\
\langle{\bar{\psi}_l,\bar{\psi}_m\rangle}&=-\delta_{l,m}, \\
\langle{\psi_l,\bar{\psi}_m\rangle}&=\langle{\bar{\psi}_l,\psi_m\rangle}=0.
\end{align}
\end{subequations}
Since the eigenvalues $\lambda_j$, $\bar{\lambda}_j$ $(j=1,\ldots,n)$ are all distinct, the vectors $\psi_j$,$\bar{\psi}_j$ $(j=1,\ldots,n)$ form a basis for $\mathbb{C}^{2n}$.  Using this fact, together with the conditions (\ref{orthonormal}), it follows that any $\zeta\in\mathbb{R}^{2n}$ may be written uniquely as:
\begin{align}
\zeta=2\mathcal{R}e\sum_{k=1}^n\langle{\zeta,\psi_k\rangle}\psi_k. \label{eigenexpansion}
\end{align}

Consider the set of quadratic functions $f_k$ given for $\zeta\in\mathbb{R}^{2n}$ by:
\begin{equation}
f_k(\zeta)=\left|\langle{\zeta,\psi_k\rangle}\right|^2\quad\quad (k=1,\ldots,n). \label{lininvariants}
\end{equation}
Then each $f_k$ is invariant under the linear map since:
\begin{equation}
f_k(R\zeta)=\left|\langle{R\zeta,\psi_k\rangle}\right|^2=\left|\langle{\zeta,R^{-1}\psi_k\rangle}\right|^2=f_k(\zeta).
\end{equation}
To obtain the second equality, we used the symplectic condition $R^TJR=J$, and to obtain the third equality, we used the facts that $R^{-1}\psi_k=\lambda_k^{-1}\psi_k$ and $|\lambda_k^{-1}|=1$, which follow from (\ref{Rconditions}).

Using (\ref{lininvariants}), one may verify that the Jacobian matrix $Df_k(\zeta)$ at the point $\zeta\in \mathbb{R}^{2n}$ acts on vectors $v$ to give:
\begin{equation}
Df_k(\zeta)v=2\mathcal{R}e\langle{\zeta,\psi_k\rangle}\langle{\psi_k,v\rangle},\quad v\in\mathbb{R}^{2n}. \label{dfdef}
\end{equation}
Likewise, the Jacobian matrix of the momentum mapping $D\mathcal{F}(\zeta)$ at any point $\zeta\in\mathbb{R}^{2n}$ becomes:
\begin{equation}
D\mathcal{F}(\zeta)=
\begin{pmatrix}
Df_1(\zeta) \\
\vdots \\
Df_n(\zeta)
\end{pmatrix}.\label{linmomentum}
\end{equation}
Using (\ref{linmomentum}), the Poisson bracket condition (\ref{PB}) takes the form:
\begin{equation}
(Df_j)J(Df_k)^T=0,\quad j,k=1,\ldots,n
\end{equation}
where we have suppressed the dependence on $\zeta$.  This follows from the orthogonality conditions (\ref{orthonormal}), using (\ref{dfdef}).

Define a $2n\times n$ matrix $B$ by:
\begin{equation}
B=\begin{pmatrix}
b_1 & \cdots & b_n
\end{pmatrix},\label{bmap}
\end{equation}
where the $b_k$ are real $2n$-vectors given by:
\begin{equation}
b_k=\mathcal{R}e\left(\psi_k/\langle{\zeta,\psi_k\rangle}\right),
\end{equation}
which are defined, provided that $f_k(\zeta)\neq 0$.
Then it follows from (\ref{linmomentum}) and (\ref{bmap}) that
\begin{equation}
[D\mathcal{F}(\zeta)B]_{jk}=Df_j(\zeta)b_k=2\mathcal{R}e\langle{\zeta,\psi_j\rangle}\langle{\psi_j,b_k\rangle},
\end{equation}
where in the last equality we used (\ref{dfdef}).  However,
\begin{equation}
\langle{\psi_j,b_k\rangle}=\frac{1}{2}\left(\frac{\langle{\psi_j,\psi_k\rangle}}{\langle{\zeta,\psi_k\rangle}}+\frac{\langle{\psi_j,\overline{\psi}_k\rangle}}{{\langle{\psi_k,\zeta\rangle}}}\right)=\frac{\delta_{jk}}{2\langle{\zeta,\psi_k\rangle}}, \label{bidentity}
\end{equation}
by the orthonormality conditions, so that
\begin{equation}
[D\mathcal{F}(\zeta)B]_{jk}=\delta_{jk},
\end{equation}
and $B$ is a right matrix inverse of $D\mathcal{F}(\zeta)$.  This shows that $\operatorname{rank}(D\mathcal{F}(\zeta))=n$, provided $f_k(\zeta)\neq 0$ for all $k=1,\ldots,n$.

We now examine the regular level sets of the momentum mapping $\mathcal{F}$, which take the form:
\begin{equation}
M_c=\{\zeta\in\mathbb{R}^{2n}:f_k(\zeta)=c_k,k=1,\ldots,n\},
\end{equation}
where $c_k\neq 0$ for all $k$.
Note that by (\ref{lininvariants}) we have
\begin{equation}
f_k(\zeta)=c_k\Leftrightarrow \langle{\zeta,\psi_k\rangle}=\sqrt{c_k}e^{i t_k},
\end{equation}
for some real phase angle $t_k$.  It follows from (\ref{eigenexpansion}) that:
\begin{equation}
\zeta\in M_c\Leftrightarrow \zeta=2\mathcal{R}e\sum_{k=1}^n\sqrt{c_k}e^{i t_k}\psi_k, \label{Mcdef}
\end{equation}
for some real $t_1,\ldots,t_n$.  Given a point $\zeta\in M_c$, applying the map $R$ gives:
\begin{equation}
R\zeta=2\mathcal{R}e\sum_{k=1}^n\sqrt{c_k}e^{it_k}R\psi_k=2\mathcal{R}e\sum_{k=1}^n\sqrt{c_k}e^{i(t_k+\phi_k)}\psi_k, \notag
\end{equation}
where in the last equality we have introduced the angle $\phi_k$ by $\lambda_k=e^{i\phi_k}$.  Define the path $\gamma:[0,1]\rightarrow M_c$ by:
\begin{equation}
\gamma(t)=2\mathcal{R}e\sum_{k=1}^n\sqrt{c_k}e^{it\phi_k}\psi_k.\label{gammapath}
\end{equation}
The tangent vector takes the form:
\begin{equation}
\gamma'(t)=2\mathcal{R}e\sum_{k=1}^ni\phi_k\sqrt{c_k}e^{it\phi_k}\psi_k. \label{gammatangent}
\end{equation}
We can now evaluate the vector quantity $S$ appearing in (\ref{invariant}).  By (\ref{bmap}), its components take the form:
\begin{equation}
S_k=\left(-\int_{\gamma}B^TJd\zeta\right)_k=-\int_0^{1}b_k^TJ\gamma'(t)dt.
\end{equation}
Using the explicit form for the tangent vector (\ref{gammatangent}) gives:
\begin{equation}
S_k=2\mathcal{R}e\sum_{j=1}^n\phi_j\sqrt{c_j}\int_0^{1}e^{it\phi_j}\langle{b_k,\psi_j\rangle}dt.
\end{equation}
Now using (\ref{bidentity}) we have:
\begin{equation}
S_k=\mathcal{R}e\phi_k\sqrt{c_k}\int_0^1\frac{e^{it\phi_k}}{\langle{\psi_k,\gamma(t)\rangle}}dt. \label{Sintegraleval}
\end{equation}
Using the explicit form of the path (\ref{gammapath}) gives:
\begin{equation}
\langle{\psi_k,\gamma(t)\rangle}=\sum_{j=1}^n\sqrt{c_j}e^{it\phi_j}\langle{\psi_k,\psi_j\rangle}+\sum_{j=1}^n\sqrt{c_j}e^{-it\phi_j}\langle{\psi_k,\overline{\psi}_j\rangle}, \notag
\end{equation}
which gives, using the conditions (\ref{orthonormal}),
\begin{equation}
\langle{\psi_k,\gamma(t)\rangle}=\sqrt{c_k}e^{it\phi_k}.
\end{equation}
Using this in (\ref{Sintegraleval}), the integral gives trivially that:
\begin{equation}
S_k=\phi_k.
\end{equation}
For the basis cycles $\gamma_k$ $(k=1,\ldots,n)$, we will take paths $\gamma_k:[0,1]\rightarrow M_c$ given by:
\begin{equation}
\gamma_k(t)=2\mathcal{R}e\sqrt{c_k}e^{i2\pi t}\psi_k, \label{cyclepath}
\end{equation}
with tangent vectors
\begin{equation}
\gamma_k'(t)=2\mathcal{R}e\sqrt{c_k}(2\pi i)e^{i2\pi t}\psi_k.
\end{equation}
Then we have:
\begin{equation}
R_{jk}=\left(-\oint_{\gamma_k}B^TJd\zeta\right)_j=-\int_0^1b_j^TJ\gamma_k'(t)dt.
\end{equation}
Using the explicit form for the tangent vector gives:
\begin{equation}
R_{jk}=2\mathcal{R}e\sqrt{c_k}(2\pi )\int_0^1e^{i2\pi t}\langle{b_j,\psi_k\rangle}dt.
\end{equation}
Now using (\ref{bidentity}) we have:
\begin{equation}
R_{jk}=\mathcal{R}e2\pi\sqrt{c_k}\delta_{jk}\int_0^1\frac{e^{i2\pi t}}{\langle{\psi_j,\gamma_k(t)\rangle}}dt. \label{Ralmost}
\end{equation}
Since this is nonzero only when $j=k$, we have in this case using the path (\ref{cyclepath}) that:
\begin{equation}
\langle{\psi_k,\gamma_k(t)\rangle}=\sqrt{c_k}e^{i2\pi t}.
\end{equation}
It follows that the integral in (\ref{Ralmost}) gives trivially that:
\begin{equation}
R_{jk}=2\pi \delta_{jk},
\end{equation}
so $R=2\pi I_{n\times n}$, and therefore (\ref{invariant}) gives the tunes:
\begin{equation}
\nu=R^{-1}S,\quad\quad \nu_k=\frac{\phi_k}{2\pi}\quad (k=1,\ldots,n),
\end{equation}
which are expressed in terms of the eigenvalues $\lambda_k=e^{i\phi_k}$, as expected \cite{Dragt}.

The freedom in (\ref{unimodular}) can be explored by making alternative choices for the paths $\gamma$ and $\gamma_k$, after noting that a general smooth path $\gamma:[0,1]\rightarrow M_c$ takes the form:
\begin{equation}
\gamma(t)=2\mathcal{R}e\sum_{j=1}^n\sqrt{c_j}e^{ig_j(t)}\psi_j, \label{gammapath2}
\end{equation}
where $g:[0,1]\rightarrow\mathbb{R}^n$ is a smooth path in $\mathbb{R}^n$.  

\section*{\label{sec:AppC}Appendix D: Special Cases in Low Dimension}

Consider a symplectic map $\mathcal{M}:\mathbb{R}^2\rightarrow\mathbb{R}^2$ given by:
\begin{equation}
    (q^f,p^f) = \mathcal{M}(q,p),
\end{equation}
together with a smooth function  $f:\mathbb{R}^2\rightarrow\mathbb{R}$ satisfying:
\begin{equation}
    f(q^f,p^f) = f(q,p), 
\end{equation}
so that $f$ is an invariant of the map $\mathcal{M}$.
Evaluating (\ref{TS0},\ref{TS}) in the special case $n=1$ shows that the rotation number of $\mathcal{M}$ on the level set $f=c$ is given by \cite{Zolkin}:
\begin{equation}
    \nu =  \frac{\int_q^{q^f}\left(\frac{\partial f}{\partial p}\right)^{-1}\,dq}
                {\oint      \left(\frac{\partial f}{\partial p}\right)^{-1}\,dq}
        =  \frac{\int_p^{p^f}\left(-\frac{\partial f}{\partial q}\right)^{-1}\,dp}
                {\oint      \left(-\frac{\partial f}{\partial q}\right)^{-1}\,dp},
\end{equation}
where each integral is taken along a path lying in the curve $f=c$, which may be parameterized by solving locally for $q$ as a function of $p$ or vice-versa.

As a special case with $n=2$, consider a symplectic map given in canonical polar coordinates as:
\begin{equation}
    (r^f,\theta^f,p_r^f,p_\theta^f) = \mathcal{M}(r,\theta,p_r,p_\theta),
\end{equation}
together with two invariants $f_1$ and $f_2$ of the form:
\begin{subequations}
\begin{align}
f_1(r,\theta,p_r,p_\theta) &= f(r,p_r,p_\theta),    \\
f_2(r,\theta,p_r,p_\theta) &= p_\theta.
\end{align}
\end{subequations}
Here $f$ is any smooth function of 3 variables.
The first invariant is independent of the angle coordinate, while the second invariant is just the angular momentum.
Choose $\gamma_1$ to be a closed curve in the invariant level set $(f_1,f_2)=(c_1,c_2)$ obtained after setting $\theta=$const.  This curve can be parameterized by solving locally for $r$ as a function of $p_r$ or vice-versa.  Choose $\gamma_2$ to be a closed curve in the same invariant level set obtained after setting $r=$const, allowing $\theta$ to vary from 0 to 2$\pi$.

Evaluating (\ref{TS0},\ref{TS}) shows that the rotation vector $\nu=(\nu_r,\nu_{\theta})$ can be written in terms of tunes associated with radial and angular motion as:
\begin{subequations}
\begin{align}
\nu_r &=
    \frac{\int_r^{r^f}\left(\frac{\partial f}{\partial p_r}\right)^{-1}\,dr}
    {\oint      \left(\frac{\partial f}{\partial p_r}\right)^{-1}\,dr} =  \frac{\int_{p_r}^{p_r^f}\left(\frac{\partial f}{\partial r}\right)^{-1}\,dp_r}
    {\oint      \left(\frac{\partial f}{\partial r}\right)^{-1}\,dp_r},       \\
\nu_\theta &=
    \nu_r\frac{\Delta_\theta}{2\,\pi} -
        \frac{\Delta_\theta'}{2\,\pi} +
        \frac{\delta\theta}{2\,\pi},
\end{align}
\end{subequations}
where the integrals are taken over all or part of the path $\gamma_1$ and:
\begin{align}
\Delta_\theta'  &= \int_r^{r^f}\frac{\partial f}{\partial p_\theta}
        \left(\frac{\partial f}{\partial p_r}\right)^{-1}\,dr
                = \int_{p_r}^{p_r^f}\frac{\partial f}{\partial p_\theta}
        \left(-\frac{\partial f}{\partial r}\right)^{-1}\,dp_r,  \notag          \\
\Delta_\theta   &= \oint      \frac{\partial f}{\partial p_\theta}
        \left(\frac{\partial f}{\partial p_r}\right)^{-1}\,dr
                = \oint      \frac{\partial f}{\partial p_\theta}
        \left(-\frac{\partial f}{\partial r}\right)^{-1}\,dp_r,   \notag         \\
\delta \theta  &=  \theta^f - \theta.
\end{align}

\end{document}